\documentclass[sigconf]{acmart}

\usepackage{subfigure}
\usepackage{graphicx}
\usepackage{amsmath}
\usepackage{bm}
\usepackage{url}
\usepackage{stmaryrd}
\usepackage{balance}
\usepackage{amssymb}
\usepackage{pgfplots}
\usepackage{pifont}
\usepackage{epsfig}
\usepackage{booktabs}
\usepackage{pgffor}
\usepackage{tikz}
\usepackage{multirow}
\usepackage{amsmath}

\makeatletter
\newcommand\xleftrightarrow[2][]{%
  \ext@arrow 9999{\longleftrightarrowfill@}{#1}{#2}}
\newcommand\longleftrightarrowfill@{%
  \arrowfill@\leftarrow\relbar\rightarrow}
\makeatother

\newcommand{\mb}{\mathbf}

\newcommand{\our}{\textsc{Humor}}

\newcommand{\ouresn}{\textsc{Humor-esn}}
\newcommand{\infesn}{\textsc{Inf-esn}}
\newcommand{\cutesn}{\textsc{Cut-esn}}
\newcommand{\kmeansesn}{\textsc{Kmeans-esn}}
\newcommand{\ourchart}{\textsc{Humor-chart}}
\newcommand{\infchart}{\textsc{Inf-chart}}
\newcommand{\cutchart}{\textsc{Cut-chart}}
\newcommand{\kmeanschart}{\textsc{Kmeans-chart}}

\newcommand{\problem}{\textsc{BL-Ecd}}

\usepackage{algorithm}
\usepackage{algorithmic}

\begin{document}

\fancyhead{}

\title{BL-ECD: Broad Learning based Enterprise Community Detection via Hierarchical Structure Fusion}





\author{Jiawei Zhang$^1$, Limeng Cui$^2$, Philip S. Yu$^{3, 4}$, Yuanhua Lv$^5$}
\affiliation{%
  $^1$IFM Lab, Department of Computer Science, Florida State University, FL, USA \\
  {$^2$School of Computer and Control Engineering, University of Chinese Academy of Sciences, Beijing, China}\\
  {$^3$University of Illinois at Chicago, Chicago, IL, USA}\\
  $^4$Shanghai Institute for Advanced Communication and Data Science, Fudan University, Shanghai, China\\
  $^5$Microsoft, Sunnyvale, CA, USA
}
\email{jzhang@cs.fsu.edu, lmcui932@163.com, psyu@cs.uic.edu, yuanhual@microsoft.com}

\begin{abstract}
Employees in companies can be divided into different communities, and those who frequently socialize with each other will be treated as close friends and are grouped in the same community. In the enterprise context, a large amount of information about the employees is available in both (1) offline company internal sources and (2) online enterprise social networks (ESNs). Each of the information sources also contain multiple categories of employees' socialization activities at the same time. In this paper, we propose to detect the social communities of the employees in companies based on the broad learning setting with both these online and offline information sources simultaneously, and the problem is formally called the ``Broad Learning based Enterprise Community Detection'' ({\problem}) problem. To address the problem, a novel broad learning based community detection framework named ``\underline{H}eterogeneo\underline{U}s \underline{M}ulti-s\underline{O}urce Cluste\underline{R}ing'' ({\our}) is introduced in this paper. Based on the various \textit{enterprise social intimacy} measures introduced in this paper, {\our} detects a set of \textit{micro community structures} of the employees based on each of the socialization activities respectively. To obtain the (globally) consistent community structure of employees in the company, {\our} further fuses these \textit{micro community structures} via two broad learning phases: (1) \textit{intra-fusion} of \textit{micro community structures} to obtain the online and offline (locally) consistent communities respectively, and (2) \textit{inter-fusion} of the online and offline communities to achieve the (globally) consistent community structure of employees. Extensive experiments conducted on real-world enterprise datasets demonstrate our method can perform very well in addressing the {\problem} problem.
\end{abstract}



\maketitle

\section{Introduction}\label{sec:introduction}


People in social organizations (e.g., schools, companies, and even countries) are usually involved in different social communities, where individuals who frequently socialize with each other will belong to the same community, while those who rarely interact with each other will be partitioned into different communities. Meanwhile, detecting the social communities of people within social organizations is formally called the \textit{community detection} problem \cite{ZY15, ZY15-2, ZYL17}. Depending on the specific application settings, the community detection problem can be studied in social organizations of various granularities. In this paper, we will take the ``company'' as an example \cite{ZYL15, ZLY15, HXZZGY16, ZYLZ16, ZYL17, ZYL17-2}, and propose to analyze the community structures formed by employees within companies.

Workplace \cite{ZYL15, ZLY15, ZYLZ16, ZYL17, ZYL17-2} is actually an easily neglected yet important social occasion for effective communication and interactions among people in our social life. Based on the professional context in enterprises, various company internal information about the employees is available, which can provide crucial signals about the community structures formed by the employees. For instance, employees in companies are usually divided into different hierarchies, where people in management are at higher levels, while the regular employees are at lower levels. The whole management structure of the employees in a company can be characterized with the \textit{enterprise organizational chart} visually \cite{ZYL15, ZLY15, ZYLZ16, ZYL17, ZYL17-2}. Generally, employees who are close in the company \textit{organizational chart} (e.g., colleagues in the same group) tend to be more familiar with each other and are highly likely to be in the same community. What's more, employees in companies all have their job titles, which describe their duties, responsibilities, and relative ranks in the company. Employees who have common job titles may need to collaborate to carry out projects together, and are more likely to be in the same community. In addition, for large international companies with branches all around the world, from the company we can know the employees' workplace information and their office locations as well. Employees who work at the same place will have more chance to meet in the real world and are more likely to be in the same community, especially compared with those working in different countries. Therefore, the company internal information provides very important clues for us to understand employees' social community structures within companies.

Meanwhile, nowadays, to facilitate the communication and collaboration among employees within companies, a new family of online social networks has been adopted in many companies, which are called the ``\textit{enterprise social networks}'' (ESNs) \cite{ZYL15, ZLY15, ZYLZ16, ZYL17, ZYL17-2}. A representative example of online ESNs is Yammer. Over 500,000 businesses around the world are now using Yammer, including 85\% of the Fortune 500. ESNs can provide employees with various types of professional services to help them deal with daily work issues, e.g., identification of cooperators with required expertise; document sharing among the team members; and instant notifications of some important internal events (like management board changes, and newly released products). What's more, in these online ESNs, employees can also perform various personal and casual social activities, like (1) making new friends online, (2) chatting with other employees on various personal and work-related topics, (3) socializing with other employees and writing comments on their posts, as well as (4) creating/joining in personal interest groups. Generally, employees who socialize with each other more frequently in online ESNs are more close and will be partitioned into the same communities. Therefore, the personal information about employees' in online ESNs is an important complement of their company internal professional information, which together will provide a more comprehensive understanding about employees' community structures in companies. 

\noindent \textbf{Problem Studied}: In this paper, we propose to study the social community structures of employees in companies based on the broad learning setting with both the company internal professional information and the personal information available in the online ESNs. Formally, the problem is named as the {\problem} (\underline{B}road \underline{L}earning based \underline{E}nterprise \underline{C}ommunity \underline{D}etection) problem. 

Here, ``\textbf{Broad Learning}'' is a new type of learning task, which focuses on fusing multiple large-scale information sources of diverse varieties together and carrying out synergistic data mining tasks across these fused sources in one unified analytic \cite{ZY15-4, ZCZCYH17, ZYZ14, ZY15, ZYL15, ZLY15, ZYLZ16, ZYL17, ZYL17-2}. In the real world, on the same information entities, e.g., social media users \cite{ZY15-4, ZCZCYH17, ZYZ14, ZY15}, movie knowledge library entries \cite{ZZHWZZY17} and employees in companies \cite{ZYL15, ZLY15, ZYLZ16, ZYL17, ZYL17-2}, a large amount of information can actually be collected from various sources. These sources are usually of different varieties, like Foursquare vs Twitter \cite{ZY15-4, ZCZCYH17, ZYZ14, ZY15}, IMDB vs Douban Movie sites \cite{ZZHWZZY17}, Yammer vs company organizational chart \cite{ZYL15, ZLY15, ZYLZ16, ZYL17, ZYL17-2}. Each information source provides a specific signature of the same entity from a unique underlying aspect. Effective fusion of these different information sources provides an opportunity for researchers and practitioners to understand the entities more comprehensively, which renders ``\textbf{Broad Learning}'' an extremely important learning task. Fusing and mining multiple information sources of large volumes and diverse varieties are a fundamental problem in big data studies. ``\textbf{Broad Learning}'' investigates the principles, methodologies and algorithms for synergistic knowledge discovery across multiple aligned information sources, and evaluates the corresponding benefits. Great challenges exist in ``\textbf{Broad Learning}'' for the effective fusion of relevant knowledge across different aligned information sources depends upon not only the relatedness of these information sources, but also the target application problems. ``\textbf{Broad Learning}'' aims at developing general methodologies, which will be shown to work for a diverse set of applications, while the specific parameter settings can be learned for each application from the training data.



The {\problem} problem is an important problem, which can be the prerequisite for many concrete applications in real-world companies \cite{ZYL15, ZLY15, HXZZGY16, ZYLZ16, ZYL17, ZYL17-2}, like \textit{team formation for company internal projects} \cite{ZYL17-2} and \textit{friend recommendations in ESNs} \cite{ZLY15}. Generally, employees who are in the same communities will know each other much better than those in other communities and the \textit{communication costs} among them will be relatively lower, which is a very important factor in \textit{team formation} problem \cite{ZYL17-2}. Meanwhile, for employees who are in the same communities, we can recommend them as potential friends to each other in the online ESNs, and they are more likely to accept the recommendations \cite{ZLY15}.

Besides its importance, the {\problem} problem is a new problem, which has never been studied before (a short version of this paper is available at \cite{ZYL17}). To the best of our knowledge, we are the first to study the community detection problem in the enterprise context. The {\problem} problem is different from existing works on community detection problems. The enterprise context generates lots of unique information sources that are rare in regular social occasions, like the the tree-structured diagram, and ESNs, which renders the existing clustering and community detection methods cannot be applied to address the {\problem} problem directly.

In addition, the {\problem} problem is very challenging to address due to the following reasons:
\begin{itemize}

\item \textit{Online ESNs based Community Extraction}: Different types of social interactions about the employees in online ESNs, e.g., friendship, interest groups memberships, posts and employee generated contents, provide important information about the employees closeness in the online world. A clear definition about the \textit{enterprise social intimacy} and the \textit{social community structure} of the employees based on the information in ESNs is desired for addressing the {\problem} problem.

\item \textit{Internal Information based Community Extraction}: Various categories of internal company information, e.g., the \textit{company organizational chart}, \textit{employees' job titles}, and \textit{employees' working locations}, all can indicate the closeness among the employees in the company from different aspects. A formal definition of the \textit{enterprise social intimacy} and the \textit{social community structure} of the employees can be the prerequisite for utilizing the company internal information to address the {\problem} problem.

\item \textit{Community Structure Fusion}: Distinct employees' community structures can be detected with these various types of information about employees in both online ENS and company internal sources respectively. An effective fusion of these information sources can provide us with a more comprehensive understanding about employees' social communities in companies, which is still an open problem so far.

\end{itemize}

To address the above challenges, a new enterprise social community detection framework {\our} (\underline{H}eterogeneo\underline{U}s \underline{M}ulti-s\underline{O}urce Cluste\underline{R}ing) is introduced in this paper. A new concept named \textit{enterprise social intimacy} is formally defined to measure the closeness among employees in this paper. {\our} calculates the intimacy scores among employees based on the heterogeneous information available in online ESNs and offline company internal sources respectively. Based on each category of the information, {\our} can detect unique community structures involving the employees, which are called the \textit{micro enterprise communities} in this paper. Framework {\our} obtains the globally consistent enterprise community structure by fusing these detected \textit{micro enterprise communities} with two hierarchical broad learning fusion phases: (1) \textit{intra-fusion} of the \textit{micro enterprise communities} detected in online ESNs and company internal sources respectively, and (2) \textit{inter-fusion} of the enterprise community structures between online ESNs and company internal sources.

The remaining part of this paper is organized as follows. At first, we introduce several important concepts and define the {\problem} problem in Section~\ref{sec:formulation}. The framework {\our} is introduced in detail in Section~\ref{sec:method}, and evaluated in Section~\ref{sec:experiment}. Finally, we talk about the related works in Section~\ref{sec:relatedwork} and conclude this paper in Section~\ref{sec:conclusion}.

\vspace{-3pt}
\section{Problem Formulation} \label{sec:formulation}

In this section, we will first define several important concepts used in this paper, based on which, we will introduce the formulation of the {\problem} problem next.

\vspace{-3pt}
\subsection{Terminology Definition}

Enterprise social networks studied in this paper are a new type of social networks launched in the firewalls of companies especially, which can be formally defined as follows.

\noindent \textbf{Definition 1} (Enterprise Social Network (ESN)): An \textit{enterprise social network} can be represented as a heterogeneous information network $G = (\mathcal{V}, \mathcal{E})$, where $\mathcal{V}$ denotes the set of nodes, and $\mathcal{E}$ represents the complex links in the network.

In this paper, we will take Yammer as an example of online ESNs. Users in Yammer can perform various kinds of social activities, e.g., (1) follow other users, (2) create and join social groups of their interests, (3) write posts, and (4) comment on/reply/like posts written by others. Therefore, according to the definition, the Yammer network studied in this paper can be represented as $G = (\mathcal{V}, \mathcal{E})$, where $\mathcal{V} = \mathcal{U} \cup \mathcal{G} \cup \mathcal{P}$ contains the node sets of the users, social groups and posts, and $\mathcal{E} = \mathcal{E}_{u,u} \cup  \mathcal{E}_{u,g} \cup \mathcal{E}_{u,p}^w \cup  \mathcal{E}_{u,p}^c \cup \mathcal{E}_{u,p}^l$ involves the links among users, those between users and groups, and the write/comment/like links between users and posts respectively. Besides the online ESNs, a large amount of information (e.g., the organizational chart, and various other attribute information) about the employees is available inside the company, which can be formally represented as an attribute augmented organizational chart.

\noindent \textbf{Definition 2} (Attribute Augmented Organizational Chart): An \textit{attribute augmented organizational chart} can be represented as a rooted tree $T = (\mathcal{N}, \mathcal{L}, root, \mathcal{A})$, where $\mathcal{N}$ denotes the set of employees in the company, $\mathcal{L}$ represents the management links from managers to subordinates, and $root \in \mathcal{N}$ indicates the CEO of the company. Set $\mathcal{A}$ is the set of attributes attached to employees in $\mathcal{N}$. For instance, in this paper, each employee in the company is associated with both the \textit{job title} and \textit{workplace} attributes. Therefore, set $\mathcal{A}$ can be represented as $\mathcal{A} = \mathcal{A}_t \cup \mathcal{A}_l$, where job title attribute set $\mathcal{A}_t = \{A_t(u_1), A_t(u_2), \cdots, A_t(u_{|\mathcal{\mathcal{N}}|})\}$ involves the job title attribute of all the employees and employee workplace attribute set $\mathcal{A}_l = \{A_l(u_1), A_l(u_2), \cdots, A_l(u_{|\mathcal{\mathcal{N}}|})\}$ contains the workplace attribute of the employees.

Generally, the organizational chart and company internal attribute information sources involve all the employees in the company and we treat the employee node set $\mathcal{N}$ in the above definition as the complete employee set. Among all the employees, some of them may be using the online ESNs. In other words, the users joining in the online ESNs is actually a subset of the whole employees, i.e., $\mathcal{U} \subseteq \mathcal{N}$, which will be considered when conducting the \textit{inter-fusion} of community structures between online ESNs and company internal sources.

\vspace{-3pt}
\subsection{Problem Definition}

Based on the concepts defined above, we can define the {\problem} problem formally in this section.

\noindent \textbf{Problem Definition} (The {\problem} Problem): Given the online ESN $G$ and the offline attribute augmented organizational chat $T$, the {\problem} problem aims at inferring the community structure of employees in the company. More specifically, we aim at partition the employee set $\mathcal{N}$ in the company into $K$ disjoint social communities $\mathcal{C} = \{C_1, C_2, \cdots, C_K\}$. In this paper, we don't consider that case that each employee is involved in multiple communities. Therefore, for these detected community structures we have $C_i \cap C_j = \emptyset, \forall i, j \in \{1, 2, \cdots, K\}, i \neq j$ and $\bigcup_i C_i = \mathcal{N}$. Generally, the employees grouped in each community (e.g., $C_i$) tend to interact with each other more frequently and have larger closeness scores with each other compared with those in other communities (i.e., $\mathcal{C} \setminus C_i$).

\section{Proposed Methods}\label{sec:method}

In this section, we will talk about the enterprise community detection framework {\our} in great detail. At first, we will introduce the definition of \textit{ESNs based enterprise social intimacy}  measure and calculate the intimacy scores among employees based on the different types of social interaction information in online ESNs in Section~\ref{subsec:esn_intimacy}. With these calculated intimacy scores, a set of \textit{micro enterprise communities} can be extracted, which will be fused with the \textit{intra-fusion} step by {\our} in Section~\ref{subsec:esn_fusion}. Similarly, in Section~\ref{subsec:chart_intimacy}, we will calculate the \textit{company internal information based intimacy} scores among employees based on the company internal information, and fuse the detected \textit{micro enterprise communities}  with the \textit{intra-fusion} step in Section~\ref{subsec:chart_fusion}. Finally, to obtain the globally consistent enterprise community structure of the employees, we will formulate the \textit{inter-fusion} step of the {\our} framework as a joint optimization problem, where will be introduced in Section~\ref{subsec:inter_fusion}.

\subsection{ESNs based Community Detection}\label{subsec:esn}

In this part, we will introduce the \textit{enterprise intimacy} concept based on ESNs. With the various categories of social information in online ESNs, we can calculate different types of enterprise intimacy scores among employees and detect the corresponding micro enterprise communities respectively.


\subsubsection{Enterprise Social Intimacy in Online ESN}\label{subsec:esn_intimacy}

Generally, social intimacy is a measure of how people closely interact with each other. At an individual level, specifically, the enterprise social intimacy involves the quality and number of connections one has with other employees in the company.

\noindent \textbf{Definition 3} (ESNs based Enterprise Social Intimacy): Based on the social interactions information available in the online \textit{enterprise social network} $G = (\mathcal{V}, \mathcal{E})$, the \textit{ESNs based enterprise social intimacy} between employees $u$ and $v$ denotes how close $u$ and $v$ are in the online ESNs $G$, which can be represented as $EI^{g}(u, v)$ in this paper.

In this section, we will mis-use ``employees'' and ``users'' to represent the individuals involved in online ESNs. As introduced in the previous section, employees in online ESNs can perform various types of social activities, based on which, different enterprise social intimacy measures can be calculated. 

\noindent \textbf{Social Connection based Enterprise Social Intimacy}

According to the definition introduced in Section~\ref{sec:formulation}, the set of social connections among the employees in online ESN $G$ can be represented as set $\mathcal{E}_{u,u}$. The connections among employees $(u, v) \in \mathcal{E}_{u,u}$ is a directed link, which denotes that employee $u$ follows $v$ in network $G$. Since the \textit{enterprise intimacy} concept is symmetrical, when calculating the intimacy scores among employees based on the social connection information, we will neglect the directions of the social links in this paper. Formally, for a certain employee user $u \in \mathcal{U}$, we can represent the set of neighbor employees that are connected with $u$ as set $\Gamma^g_s(u) = \{v | v \in \mathcal{U}, (u, v) \in \mathcal{E}_{u,u} \lor (v, u) \in \mathcal{E}_{u,u}\}$, where the subscript $s$ denotes that it is based on the \underline{s}ocial connection information and the superscript $g$ indicates it is extracted from the online ESN $G$. Similarly, we can also represent the set of neighbors of employee $v$ as set $\Gamma^g_s(v)$. Furthermore, the common neighbors shared by $u$ and $v$ can be denotes as set $\Gamma^g_s(u) \cap \Gamma^g_s(v)$, which is a strong indicator about the enterprise social intimacy between employees $u$ and $v$, and close employees would share more common neighbors. 

However, simply counting the common neighbors may suffer from some problems: employees $u$ and $v$ can be connected by a large number of common neighbors merely because of the large social degrees of either $u$, $v$ or the intermediate nodes between them, but the affinity between $u$ and $v$ is not high. To address this problem, we propose to weight the degrees of employees $u$ and $v$ to penalize those popular ones in the closeness calculation. To achieve such a goal, we propose the following pointwise mutual information based social intimacy between employees $u$ and $v$:
\begin{align*}
EI^{g}_s(u, v) = \frac{|\Gamma^g_s(u) \cap \Gamma^g_s(v)|}{|\mathcal{U}|} \log \frac{\frac{|\Gamma^g_s(u) \cap \Gamma^g_s(v)|}{|\mathcal{U}|}}{\frac{|\Gamma^g_s(u)|}{|\mathcal{U}|} \cdot \frac{|\Gamma^g_s(v)|}{|\mathcal{U}|}}.
\end{align*}

\noindent \textbf{Group Membership based Enterprise Social Intimacy}

Besides social connections, employee users can also create/join their personal interest groups in ESNs. Intuitively, sharing more common interest groups often suggests that two employees have more common interests and are closer to each other. Based on the group membership information in online ESNs, we propose to calculate the group participation activity based enterprise social intimacy among employees. As introduced before, in the online ESN $G$, the employees' group membership links can be represented as set $\mathcal{E}_{u, g}$, and the set of groups that employee $u$ is involved in can be represented as $\Gamma^g_{g}(u) \subseteq \mathcal{G}$. Similarly, we can represent group set for $v$ as $\Gamma^g_{g}(v) \subseteq \mathcal{G}$, and the groups shared by $u$ and $v$ can be represented as $\Gamma^g_{g}(u) \cap \Gamma^g_{g}(v)$.

However, in online ESNs, different groups may have different discrimination power, where larger groups involving a large number of members can hardly show the intimacy among the participants. More specifically, we can represent the set of employees involved in a group $g$ in the online ESN as set $\Gamma^g_g(g) \subset \mathcal{U}$. For any two different social groups $g_1$ and $g_2$ both containing employees $u$ and $v$, group $g_1$ will have a greater discrimination power than $g_2$ if $\Gamma^g_g(g_1) < \Gamma^g_g(g_2)$. In order to capture the discrimination of different groups, we propose to apply the ``Inverse Membership Frequency'' (IMF) introduced in \cite{ZLY15}, inspired by the IDF measure \cite{J97} widely used in information retrieval, and the IMF of group $g$ can be represented as $\mbox{IMF}(g) = \log \frac{|\mathcal{U}|}{|\Gamma^g_g(g)|}$. 

Furthermore, the ESN group membership based enterprise social intimacy between any two employees $u$ and $v$ can be represented as the sum of the shared groups' IMF measures\begingroup\makeatletter\def\f@size{9}\check@mathfonts
\begin{align*}
{EI}^g_g(u, v) &= \sum_{g \in \Gamma^g_{g}(u) \cap \Gamma^g_{g}(v)} \mbox{IMF}(g) = \sum_{g \in \Gamma^g_{g}(u) \cap \Gamma^g_{g}(v)} \log \frac{|\mathcal{U}|}{|\Gamma^g_g(g)|}.
\end{align*}\endgroup

\noindent \textbf{Text Content based Enterprise Social Intimacy}

In addition to the social connection and group membership information, employees in online ESNs can also write/reply/comment on/like posts, which will generate lots of text content related information. In this paper, we will not go deep into dealing with the text words, and simply calculate the enterprise social intimacy based on the correlated posts shared by different employees. For instance, we can represent the sets of posts that employee $u$ and $v$ have written/replied/commented on/liked as $\Gamma^g_{p}(u) \subseteq \mathcal{P}$ and $\Gamma^g_{p}(v) \subseteq \mathcal{P}$ respectively. The set of shared posts that $u$ and $v$ both correlate to can be represented as $\Gamma^g_{p}(u) \cap \Gamma^g_{p}(v)$. Generally, the more interactions employees have via the posts online, the closer they would be, but the total number of posts employees are correlated to (i.e., the degrees of $u$ and $v$) also need to be considered carefully. Therefore, based on the user generated content (i.e., post) information in online ESNs, can represent the enterprise social intimacy between $u$ and $v$ as the Jaccard's Coefficient (JC) of the post sets about employees $u$ and $v$:
\begin{align*}
{EI}^g_p(u, v) &= \mbox{JC}(\Gamma^g_{p}(u), \Gamma^g_{p}(v)) = \frac{|\Gamma^g_{p}(u) \cap \Gamma^g_{p}(v)|}{|\Gamma^g_{p}(u) \cup\Gamma^g_{p}(v) |}.
\end{align*}

\subsubsection{Intra-Fusion of ESNs based Enterprise Community}\label{subsec:esn_fusion}

\noindent \textbf{Micro Community Structure Detection in ESNs}

Based on each enterprise social intimacy measure introduced in the previous section, a concrete enterprise social intimacy matrix can be constructed. For instance, according to the social connection based enterprise social intimacy, we can define the enterprise social intimacy matrix $\mb{A}^{g}_s \in \mathbb{R}^{|\mathcal{U}| \times |\mathcal{U}|}$, where entry ${A}^{g}_s(i,j) = EI^g_s(u_i, u_j)$, $u_i, u_j \in \mathcal{U}$ denotes the social connection information based enterprise social intimacy between employees $u_i$ and $u_j$. In a similar way, we can represent the enterprise social intimacy scores among employees based on the group participation and user generated content information as matrices $\mb{A}^{g}_g, \mb{A}^{g}_p \in \mathbb{R}^{|\mathcal{U}| \times |\mathcal{U}|}$ respectively.

Based on each of the enterprise social intimacy matrix (e.g., $\mb{A}^{g}_s$), various methods can be applied to detect the community structure of the employees. In this paper, we propose to use the non-negative matrix factorization (NMF) method to partition the intimacy matrix $\mb{A}^{g}_s$ due to its outstanding performance and wide applications in clustering problems \cite{YL13}. The NMF method aims at minimizing the following objective equation to infer the community structure hidden factor matrix $\mb{U}$:
\begin{align*}
J_C(\mb{A}^{g}_s) = \left\| \mb{A}^{g}_s - \mb{U}_s\mb{U}_s^\top \right\|^2_F.
\end{align*}

Here, the entry $U_s(i, j)$ in the hidden factor matrix represents the confidence of employee $u_i \in \mathcal{U}$ belonging to the $j_{th}$ community $C_j \in \mathcal{C}$. Formally, the enterprise social community structure described by matrix $\mb{U}_s$ is called the \textit{micro enterprise community structure} in this paper.

Similarly, we can define the objective equations $J_C(\mb{A}^{g}_g)$ and $J_C(\mb{A}^{g}_p)$ of the enterprise social intimacy matrix based on the group membership and user generated content information. The corresponding hidden factor matrices $\mb{U}_g$ and $\mb{U}_p$ can be obtained by minimizing the following two objective equations\begingroup\makeatletter\def\f@size{9}\check@mathfonts
\begin{align*}
J_C(\mb{A}^{g}_g) &= \left\| \mb{A}^{g}_g - \mb{U}_g\mb{U}_g^\top \right\|^2_F,\ J_C(\mb{A}^{g}_p) = \left\| \mb{A}^{g}_p - \mb{U}_p\mb{U}_p^\top \right\|^2_F.
\end{align*}\endgroup

\noindent \textbf{Intra-Fusion of Micro Community Structures in ESNs}

The detected hidden factor matrices $\mb{U}_s, \mb{U}_g, \mb{U}_p$ characterize the community structures of the employee users in ESNs from different aspects, which can be different from each other. Meanwhile, these hidden factor matrices are all about community structures of the same set of employees in the online ESNs information source, which should be consistent with the real-world community structure of the employees. To achieve such a goal, in this paper, an \textit{consistent hidden factor matrix} $\mb{U} \in \mathbb{R}^{|\mathcal{U}| \times K}$ is introduced to represent the consistent community structure of the employees in online ESNs. To infer $\mb{U}$, we add a set of extra regularization terms to minimize the differences of community structure described by $\mb{U}$ from those described by $\mb{U}_s, \mb{U}_g, \mb{U}_p$ respectively:\begingroup\makeatletter\def\f@size{9}\check@mathfonts
\begin{align*}
J_R(\mb{A}^{g}_s, \mb{A}^{g}_g, \mb{A}^{g}_p) = \left\| \mb{U}_s - \mb{U} \right\|^2_F + \left\| \mb{U}_g - \mb{U} \right\|^2_F + \left\| \mb{U}_p - \mb{U} \right\|^2_F.
\end{align*}\endgroup

Formally, the process of inferring the consistent community structure matrix $\mb{U}$ is called the \textit{intra-fusion} phrase in the {\our} framework. Based on the above remarks, the optimal consistent community structure matrix $\mb{U}^*$ which can minimize the matrix decomposition costs and regularization terms simultaneously can be obtained be resolving the following objective equation:\begingroup\makeatletter\def\f@size{9}\check@mathfonts
\begin{align*}
&\mb{U}^*=\arg_{\mb{U}} \min_{\mb{U}, \mb{U}_s, \mb{U}_g, \mb{U}_p} \sum_{i \in \{s, g, p\}}J_C(\mb{A}^{g}_i) + \alpha \cdot J_R(\mb{A}^{g}_s, \mb{A}^{g}_g, \mb{A}^{g}_p),
\end{align*}\endgroup
where $\alpha$ represents the weight of the consistency regularizer and it is assigned with value $1$ in the experiments.

\subsection{Company Internal Information based Enterprise Community Detection}\label{subsec:chart}

Besides the information available in online ESNs, employees also have a large amount of information available in the offline company information, which can also help identify the social community structures of the employees. 

\subsubsection{Internal Information based Enterprise Intimacy}\label{subsec:chart_intimacy}

In this section, we will measure the enterprise social intimacy among employees based on the company internal information, which include the organizational chart, job titles and workplaces.

\noindent \textbf{Organizational Chart based Enterprise Intimacy}

Company organizational chart is a tree structured diagram outlining the management relationships among employees, where employees are connected by the ``management'' links between managers and their subordinates. For any two given employees $u$ and $v$, an intuitive idea to represent their intimacy is using the number of required steps to walk between them along the ``management'' links in the organizational chart. Generally, employees connected by less ``management links'' (e.g., colleagues in the same group) are closer to each other in the organizational chart. Based on such an observation, we propose to follow the closeness measure ``organizational intimacy'' introduced in \cite{ZLY15} for employees $u$ and $v$ as the inverse of required number of steps to walk from $u$ to $v$ via the links in the organizational chart. 


Formally, let $\mbox{step}(u, v)$ denote the number of steps needed to walk from $u$ to $v$ along the links in the adjusted organizational chart, based on which the enterprise social intimacy $EI_c^t(u, v)$ between $u$ and $v$ can be represented as
\begin{align*}
EI_c^t(u, v) &= \frac{1}{\mbox{step}(u, v)},
\end{align*}
where the subscripts $c$ denotes that $EI_c^t(u, v)$ is based on the \underline{c}hart information, and the superscript $t$ denotes that the information is from the company internal information in $T$.

\noindent \textbf{Job Title based Enterprise Intimacy}



Employees in the enterprise have different job titles to indicate their duties and relative ranks in the companies. From the point of view of job content, employees sharing common functionality on job titles will have very similar workload and work categories, and they will have more chance to cooperate with each other. To use the job title information to calculate the enterprise social intimacy among employees in enterprises, we propose to extract the root job terms from all job titles, which can categorize what the job is actually about. For instance, the root job term of both ``Senior SDE'' and ``SDE'' are both ``SDE'', but that of ``Senior Researcher'' will be ``Researcher'' instead. Formally, for any two employees, if their job titles are of the same job categories (i.e., sharing the common root job term), then we will apply the Jaccard's Coefficient based method to calculate their enterprise intimacy; otherwise, their enterprise social intimacy will be $0$ instead. Let $\mbox{job}(u)$ and $\mbox{root}(u)$ represent the bag of word representation and the root job term of employee $u$'s job title respectively. We can represent the job title based enterprise social intimacy between employees $u$ and $v$ as:
\begin{align*}
EI^t_t(u, v) = I\left(\mbox{root}(u), \mbox{root}(v)\right) \frac{|\mbox{job}(u) \cap \mbox{job}(v)|}{|\mbox{job}(u) \cup \mbox{job}(v)|},
\end{align*}
where $I\left(\mbox{root}(u), \mbox{root}(v)\right)=1$ iff $\mbox{root}(u) = \mbox{root}(v)$.

Here, we need to clarify that we don't mean employees with similar job titles will be close to each other for certain. Actually, the job title provides an important signal for us to infer their potential closeness. In many companies, employees taking similar category of workload will work in the same workplace (e.g., HR people on one floor, SDEs work together at another floor), and they have more chance to meet each other. The job title together with other categories of information will provide more comprehensive knowledge about the closeness among employees at workplace.

\noindent \textbf{Workplace based Enterprise Intimacy}

Besides the organizational chart and job title information, we can also know the workplaces of employees from the company, which can help denote the geographical intimacy among employees. Specifically, employees' workplace information includes both the \textit{country} and \textit{time zone} of employees' offices. Generally, employees working in closer locations (e.g., in the same country, or the same time zone) have more chance to interact with each other. Therefore, in this paper, we propose to measure the enterprise intimacy among employees by checking whether they are in the same country, and the same time zone or not. Let $\mbox{country}(u)$ and $\mbox{time zone}(u)$ represent the country and time zone of employee $u$'s offices. The workplace based enterprise intimacy between employees $u$ and $v$ can be represented as\begingroup\makeatletter\def\f@size{9}\check@mathfonts 
\begin{align*}
EI^t_l(u, v) = \frac{1}{2}\Big(I\big(\mbox{country}(u), \mbox{country}(v)\big) + I\big(\mbox{zone}(u), \mbox{zone}(v)\big)\Big),
\end{align*}\endgroup
where $I\big(\mbox{country}(u), \mbox{country}(v)\big) = 1$ iff $\mbox{country}(u) = \mbox{country}(v)$, and similarly for function $I\big(\mbox{zone}(u), \mbox{zone}(v)\big)$.

\subsubsection{Intra-Fusion of Internal Information based Enterprise Community}\label{subsec:chart_fusion}

Based on the above enterprise social intimacy measures, we can represent the intimacy scores among employees calculated based on organizational chart, job title and workplace information in the offline enterprise as adjacency matrices $\mb{A}^t_c, \mb{A}^t_t, \mb{A}^t_l \in \mathbb{R}^{|\mathcal{N}| \times |\mathcal{N}|}$ respectively. Similarly, the non-negative matrix factorization (NMF) method can be applied to partition the intimacy matrices to obtain their corresponding community hidden factor matrices, and the decomposition cost functions are listed as follows:\begingroup\makeatletter\def\f@size{9}\check@mathfonts
\begin{align*}
J_C(\mb{A}^t_c) &= \left\| \mb{A}^t_c - \mb{V}_c\mb{V}_c^\top \right\|_F^2, J_C(\mb{A}^t_t) = \left\| \mb{A}^t_t - \mb{V}_t\mb{V}_t^\top \right\|_F^2, \\
J_C(\mb{A}^t_l) &= \left\| \mb{A}^t_l - \mb{V}_l\mb{V}_l^\top \right\|_F^2.
\end{align*}\endgroup

Meanwhile, these community hidden factor matrices $\mb{V}_c$, $\mb{V}_t$ and $\mb{V}_l$ all describe the \textit{micro enterprise community structures} of the employees in the offline enterprise from different perspectives, which should be consistent with each other as well. To define the consistency regularization term, a new consistent community hidden factor matrix $\mb{V} \in \mathbb{R}^{|\mathcal{N}| \times K}$ is introduced, and the consistency regularization term between $\mb{V}$ and $\mb{V}_c$, $\mb{V}_t$, $\mb{V}_l$ can be represented as follows:\begingroup\makeatletter\def\f@size{9}\check@mathfonts
\begin{align*}
J_R(\mb{A}^t_c, \mb{A}^t_t, \mb{A}^t_l) = \left\| \mb{V}_c - \mb{V} \right\|^2_F + \left\| \mb{V}_t - \mb{V} \right\|^2_F + \left\| \mb{V}_l - \mb{V} \right\|^2_F.
\end{align*}\endgroup

Based on the above remarks and the \textit{intra-fusion} strategy, the optimal consistent community structure matrix $\mb{V}^*$ of employees in the offline enterprise can be obtained by resolving the following objective equation:\begingroup\makeatletter\def\f@size{9}\check@mathfonts
\begin{align*}
&\mb{V}^* =\arg_{\mb{V}} \min_{\mb{V}, \mb{V}_c, \mb{V}_t, \mb{V}_l} \sum_{i \in \{c, t, l\}}J_C(\mb{A}^t_i) + \alpha \cdot J_R(\mb{A}^t_c, \mb{A}^t_t, \mb{A}^t_l),
\end{align*}\endgroup
where the same parameter $\alpha$ (with value $1$ in the experiments) is also used to represents the weight of the consistency regularization term here.

\subsection{Inter-Fusion of Enterprise Communities}\label{subsec:inter_fusion}

Generally, the company contains the complete information about all the employees, and some of whom can get involved in the online ESNs as well. In this section, we will introduce the joint optimization objective function for detecting the communities via the \textit{inter-fusion} of information available in the online ESNs and offline company.



For the common employees shared by the online ESNs and offline company internal information sources, the detected community structures about them should be consistent with each other. To achieve such a goal, we propose to regularize the detected community structures from online ESNs and offline company internal information sources. Before introducing the regularization term, we first define the binary employee transition matrix $\mb{T} \in \mathbb{R}^{|\mathcal{U}| \times |\mathcal{N}|}$, where entry $T(i, j) = 1$ iff employee $u_i$ and $u_j$ are actually the same employee in the online ESNs and offline enterprise respectively. Based on the transitional matrix, we can map the employees as well as their hidden community matrix from the online ESN to the offline enterprise by simply multiplying it with the transition matrix. For instance, given the hidden community structure matrix $\mb{U}$ of the employees online, we can represent the mapped community structure matrix to the offline company as $\mb{T}^\top \mb{U}$. The difference of the employees' community structure mapped from the online ESNs (i.e., $\mb{T}^\top \mb{U}$) and that obtained from the offline enterprise (i.e., $\mb{V}$) can be represented as inter-source community regularization term\begingroup\makeatletter\def\f@size{9}\check@mathfonts
\begin{align*}
J_D(\mb{U}, \mb{V}) = \left\| (\mb{T}^\top\mb{U}) (\mb{T}^\top\mb{U})^\top - \mb{V}\mb{V}^\top \right\|_F^2,
\end{align*}\endgroup
where the non-zero entries in matrices $(\mb{T}^\top\mb{U}) (\mb{T}^\top\mb{U})^\top$ and $\mb{V}\mb{V}^\top$ denote the confidence scores for the corresponding employees to be in the same community based on the information available in online ESNs and offline company internal data respectively.

By adding the inter-source community regularization term, we can redefine the enterprise community detection objective equation. Formally, the process of adding the inter-source community regularization term $J_D(\mb{U}, \mb{V})$ to the objective function is named as the \textit{inter-fusion} procedure in this paper. The new objective equation can be represented as:\begingroup\makeatletter\def\f@size{9}\check@mathfonts
\begin{align*}
&\min_{\mb{U}, \mb{V}} \sum_{i \in \{s, g, p\}} J_C(\mb{A}^g_i) + \alpha \cdot J_R(\mb{A}^g_s, \mb{A}^g_g, \mb{A}^g_p)\\
& + \sum_{i \in \{c, t, l\}} J_C(\mb{A}^t_i) + \alpha \cdot J_R(\mb{A}^t_c, \mb{A}^t_t, \mb{A}^t_l) + \beta \cdot J_D(\mb{U}, \mb{V}),
\end{align*}\endgroup
where parameter $\beta$ denotes the weight of the inter-source community regularization term, whose sensitivity analysis is also available in Section~\ref{sec:experiment}. 

By replacing the cost and regularization functions with the matrix representations, we can obtain the final joint objective function involving $8$ hidden factor matrix variables, each of which can represent the community structure of the employees from different aspects. However, simultaneous inference of the optimal values for the matrix variables can be very computational hard and time consuming. 

To simplify the problem, in this paper, we propose to constrain the \textit{intra-fusion} regularization terms $J_R(\mb{A}^g_1, \mb{A}^g_2, \mb{A}^g_3)$ as well as $J_R(\mb{A}^t_c, \mb{A}^t_t, \mb{A}^t_l)$ to be both $0$, i.e., 
\begin{align*}
\mb{U}_s = \mb{U}_g = \mb{U}_p = \mb{U},\\
\mb{V}_c = \mb{V}_t = \mb{V}_l = \mb{V}.
\end{align*}

And the simplified objective function will contain two variables $\mb{U}$ and $\mb{V}$ only, which can be represented as\begingroup\makeatletter\def\f@size{9}\check@mathfonts
\begin{align*}
&\min_{\mb{U}, \mb{V}}  \left\| \mb{A}^{g}_s - \mb{U}\mb{U}^\top \right\|^2_F + \left\| \mb{A}^{g}_g - \mb{U}\mb{U}^\top \right\|^2_F + \left\| \mb{A}^{g}_p - \mb{U}\mb{U}^\top \right\|^2_F\\
&+ \left\| \mb{A}^t_c - \mb{V}\mb{V}^\top \right\|^2_F + \left\| \mb{A}^t_t - \mb{V}\mb{V}^\top \right\|^2_F + \left\| \mb{A}^t_l - \mb{V}\mb{V}^\top \right\|^2_F \\
&+ \beta \cdot \left\| (\mb{T}^\top\mb{U}) (\mb{T}^\top\mb{U})^\top - \mb{V}\mb{V}^\top \right\|_F^2.
\end{align*}\endgroup

The objective equation is not actually jointly convex and no closed-form solution exists. In this paper, we propose to solve with an alternative updating approach. We will fix one variable (e.g., $\mb{V}$) and update the other variable (e.g., $\mb{U}$) iteratively and alternatively, and such a process continues until all the variables converge.

Let $\mathcal{L}(\mb{U}, \mb{V})$ denote the objective function. In iteration $\tau$, the updating equations can be represented as
\begingroup\makeatletter\def\f@size{7}\check@mathfonts
\begin{align*}
&\mb{U}^{(\tau)} = \mb{U}^{(\tau - 1)} - \eta_1 \cdot \frac{\partial \mathcal{L}(\mb{U}^{(\tau - 1)}, \mb{V}^{(\tau - 1)})}{\partial \mb(U)}\\
&= \mb{U}^{(\tau - 1)} - 2\eta_1 \Big( 6\mb{U}^{(\tau - 1)}(\mb{U}^{(\tau - 1)})^\top\mb{U}^{(\tau - 1)} - \big(\mb{A}^{g}_s + \mb{A}^{g}_g + \mb{A}^{g}_p\big)\mb{U}^{(\tau - 1)}\\
& - \big((\mb{A}^{g}_s)^\top + (\mb{A}^{g}_g)^\top + (\mb{A}^{g}_p)^\top \big)\mb{U}^{(\tau - 1)} - 2\mb{T}\mb{V}^{(\tau - 1)}(\mb{V}^{(\tau - 1)})^\top\mb{T}^\top\mb{U}^{(\tau - 1)} \\
&+2\mb{T}\mb{T}^\top\mb{U}^{(\tau - 1)} (\mb{U}^{(\tau - 1)})^\top\mb{T}\mb{T}^\top\mb{U}^{(\tau - 1)}\Big).
\end{align*}
\begin{align*}
&\mb{V}^{(\tau)} = \mb{V}^{(\tau - 1)} - \eta_2 \cdot \frac{\partial \mathcal{L}(\mb{U}^{(\tau)}, \mb{V}^{(\tau - 1)})}{\partial \mb(V)}\\
&=\mb{V}^{(\tau - 1)} - 2\eta_2 \Big( 8\mb{V}^{(\tau - 1)}(\mb{V}^{(\tau - 1)})^\top\mb{V}^{(\tau - 1)} - \big(\mb{A}^{t}_c + \mb{A}^{t}_t + \mb{A}^{t}_l\big)\mb{V}^{(\tau - 1)}\\
&- \big((\mb{A}^{t}_c)^\top + (\mb{A}^{t}_t)^\top + (\mb{A}^{t}_l)^\top \big)\mb{V}^{(\tau - 1)} -2\mb{T}^\top\mb{U}^{(\tau)}(\mb{U}^{(\tau)})^\top\mb{T}\mb{V}^{(\tau - 1)}\Big).
\end{align*}
\endgroup
where parameters $\eta_1$ and $\eta_2$ denote the gradient descent steps in updating matrices $\mb{U}$ and $\mb{V}$ respectively. The optimal learning rates $\eta_1$ and $\eta_2$ in each iteration steps obtaining the minimum $\mathcal{L}(\mb{U}, \mb{V})$ can be represented as
\begin{align*}
\eta_1^{(\tau)} &= \arg_{\eta_1} \min \mathcal{L}(\mb{U}^{(\tau)}, \mb{V}^{(\tau)}),\\
\eta_2^{(\tau)} &= \arg_{\eta_2} \min \mathcal{L}(\mb{U}^{(\tau)}, \mb{V}^{(\tau)}).
\end{align*}
The functions can be addressed by taking derivative of $\mathcal{L}(\cdot)$ with regards to $\eta_i$ (or $\eta_2$) and make it equal to $0$, we can obtain a cubic equation involving $\eta_i$ (or $\eta_2$). Multiple roots may exist when addressing the equation and the representation of the roots is very complicated. In this paper, for simplicity, we propose to assign $\eta_i$ and $\eta_2$ with a small constant value (i.e., $0.05$ in the experiments). The above alternative updating scheme involves the multiplication of matrices. Let $n$ be the number of employees in the company and $\tau$ be the required rounds to achieve convergence, the time complexity of the alternative updating method can be represented $O(\tau n^{2.373})$, where the Optimized Coppersmith-Winograd algorithm \cite{CW90} is applied in the matrix product calculation. In addition, some pre-computation can be applied to effective reduce the real running time of the framework by storing the results of matrices $(\mb{A}^{g}_s)^\top + (\mb{A}^{g}_g)^\top + (\mb{A}^{g}_p)^\top$, $\mb{T}\mb{T}^\top$, etc., in advance.

\section{Experiments}\label{sec:experiment}


%

To test the effectiveness of the proposed framework {\our} in detecting the communities in companies, extensive experiments have been done on real-world enterprise datasets. In this section, we will first describe the datasets used in this paper, and then introduce the experiment settings in detail. Finally, we will show the experiment results together with detailed explanation and give the parameter sensitivity analysis.

\begin{figure*}[t]
\vspace{-40pt}
\centering
\subfigure[Rand]{ \label{result_truth_1}
    \begin{minipage}[l]{.45\columnwidth}
      \centering
      \includegraphics[width=\textwidth]{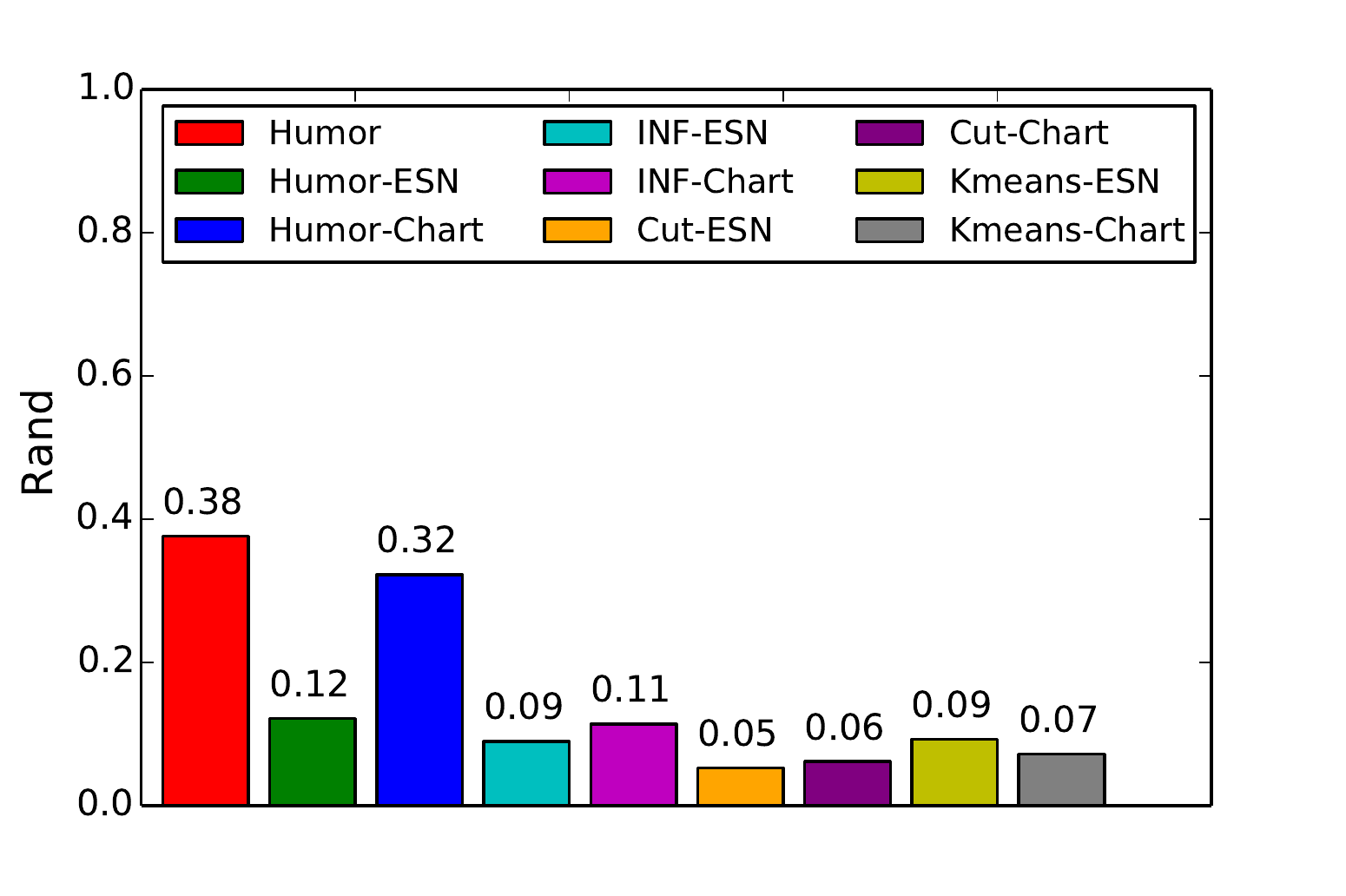}
    \end{minipage}
  }
  \subfigure[Mutual Information]{\label{result_truth_2}
    \begin{minipage}[l]{.45\columnwidth}
      \centering
      \includegraphics[width=\textwidth]{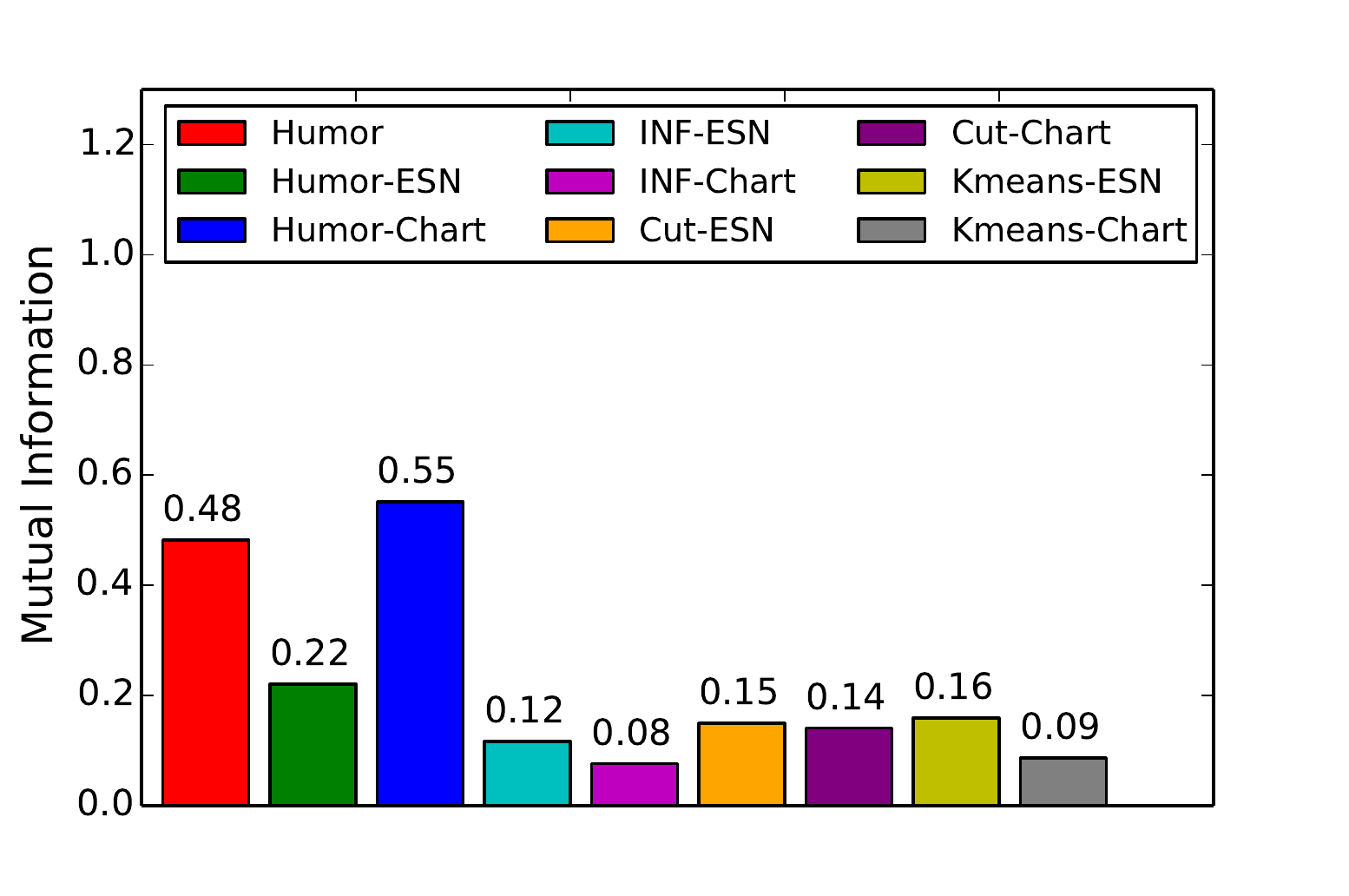}
    \end{minipage}
  }
\subfigure[Purity]{ \label{result_truth_3}
    \begin{minipage}[l]{.45\columnwidth}
      \centering
      \includegraphics[width=\textwidth]{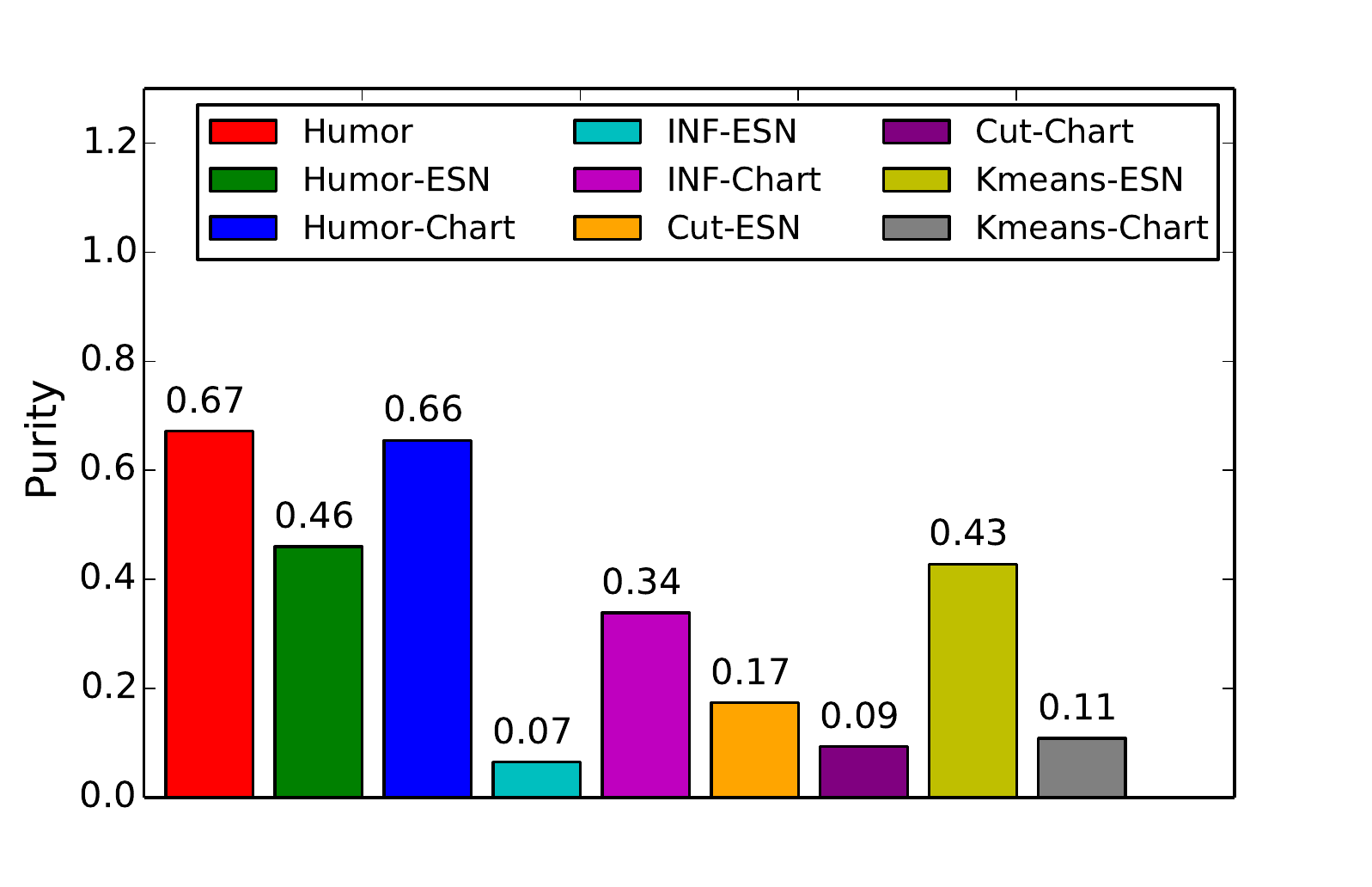}
    \end{minipage}
  }
\subfigure[Inverse Purity]{ \label{result_truth_4}
    \begin{minipage}[l]{.45\columnwidth}
      \centering
      \includegraphics[width=\textwidth]{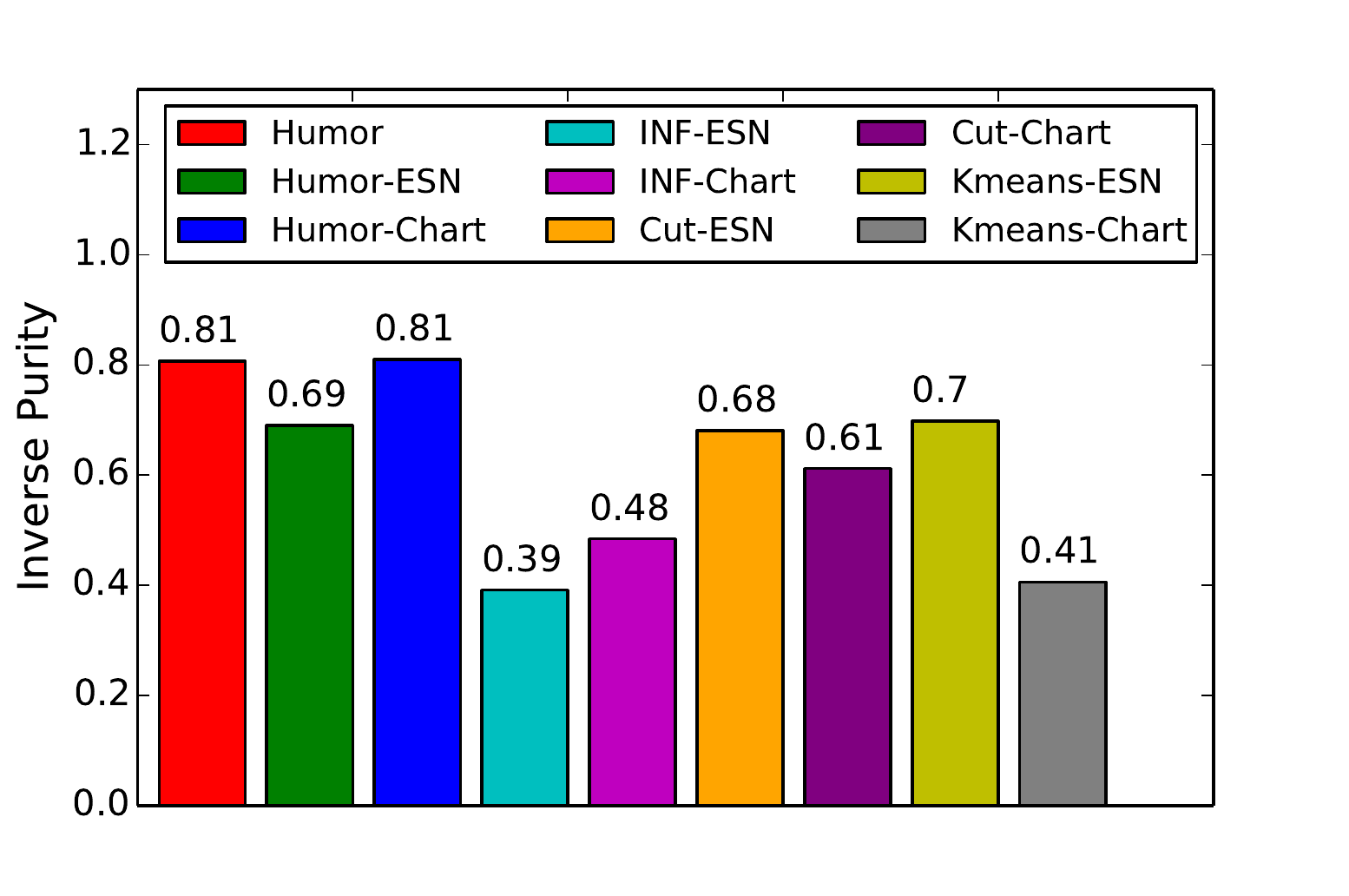}
    \end{minipage}
  }
\vspace{-15pt}
  \caption{Experiment result of comparison methods evaluated by ground truth with Rand, Mutual Information, Purity and Inverse Purity metrics.}\label{fig:result_truth}\vspace{-15pt}
\end{figure*}

\subsection{Dataset Description}

The enterprise datasets used in this paper is about the Microsoft company. We crawled all the data about the employees from Yammer as well as the complete organizational chart during June, 2014 \cite{ZYL15, ZLY15, ZYLZ16, ZYL17, ZYL17-2}. Brief descriptions about both the Yammer enterprise social network data and the Microsoft organizational chart are available as follows.\footnote{We are not able to reveal the actual numbers here and throughout the paper for commercial reasons.}

\begin{itemize}
\item \textit{Enterprise Social Network}: We crawl all the Microsoft employees' information from Yammer and obtain the complete organizational chart involving all these employees in Microsoft during June, 2014. The social network data covers all the user-generated content (such as posts, replies, topics, etc.) and social graphs (such as user-user following links, user-group memberships, user-topic following links, etc.) by then that are set to be public.

\item \textit{Organizational Chart}: Besides the online \textit{enterprise social network} and \textit{enterprise organizational chart} datasets, we can also obtain the job titles and the specific workplaces of all the employees from Microsoft, which together with the \textit{company organizational chart} is treated as the company internal information in this paper.
\end{itemize}

%

These above datasets are used as the online and offline information sources for building the models only (e.g., to calculate the intimacy scores). Meanwhile, to evaluate the proposed framework, we further crawled a dataset containing employees together their corresponding research teams from the official website of Microsoft Research (MSR)\footnote{http://research.microsoft.com/en-us/}. Here, we need to clarify that the research teams crawled from MSR are not the same as the organizational chart. Each research team denotes an on-going project carried in MSR, in which the team members may not necessary from the same departments. In the research teams, researchers with different skills and from different departments (in the organizational chart) are invited to cooperate together to carry out certain research/production projects that the team focuses on. For many research teams crawled from MSR, actually the team members can actually come from the campuses in different countries (e.g., MSR@US Redmond, MSR@US Silicon Valley, MSR@CN Beijing, MSR@UK Cambridge).

After necessary data cleaning and employee ID alignment with the \textit{enterprise social network} and \textit{organization chart} datasets, finally $142$ research teams in the MSR department. Each team in the dataset is treated as a community, and the employee-team membership is used as the ground truth of the enterprise community structure.



\subsection{Experiment Settings}

In this part, we will introduce the experiment settings, which covers the detailed experiment setup, comparison methods and the evaluation metrics.

\subsubsection{Experiment Setup}

Based on the social connection, group participation and employee generated post information available in the online ESNs, we can construct the adjacency matrices $\mb{A}^g_s$, $\mb{A}^g_g$ and $\mb{A}^g_p$ involving the all the target employees $\mathcal{U}$ to be partitioned. Meanwhile, based on the organizational chart, job titles, and specific work location information available in the company internal sources, we can construct the adjacency matrices $\mb{A}^t_c$, $\mb{A}^t_t$ and $\mb{A}^t_l$ respectively. By incorporating these constructed matrices into the objective function and obtaining the final community hidden factor matrix $\mb{V}$, we can infer the community structures of the target employee set by applying some traditional clustering models (e.g., kmeans) to matrix $\mb{V}$.

In the experiments, we will first apply the {\our} model to the small-sized MSR dataset, whose performance can be measured with the ground truth community structure, i.e., the research teams (each teams is treated as a community). In addition, we also apply the {\our} model to all whole employees in Microsoft, and the performance of {\our} is measured with some traditional evaluation metrics, which will be introduced at the end of this part.


\begin{figure*}[t]
\vspace{-40pt}
\centering
\subfigure[Density]{ \label{result_1}
    \begin{minipage}[l]{.45\columnwidth}
      \centering
      \includegraphics[width=\textwidth]{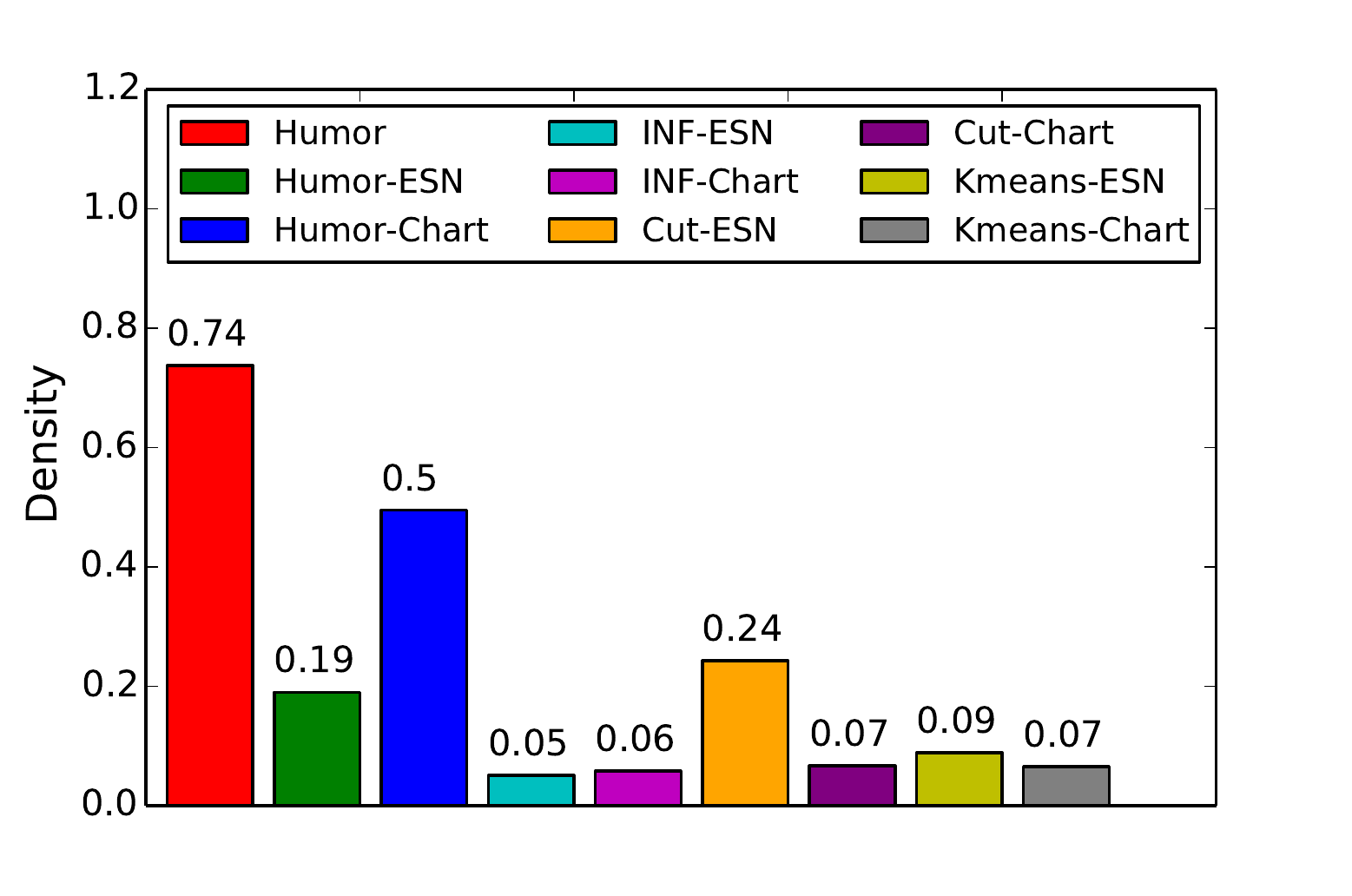}
    \end{minipage}
  }
  \subfigure[Silhouette Index]{\label{result_2}
    \begin{minipage}[l]{.45\columnwidth}
      \centering
      \includegraphics[width=\textwidth]{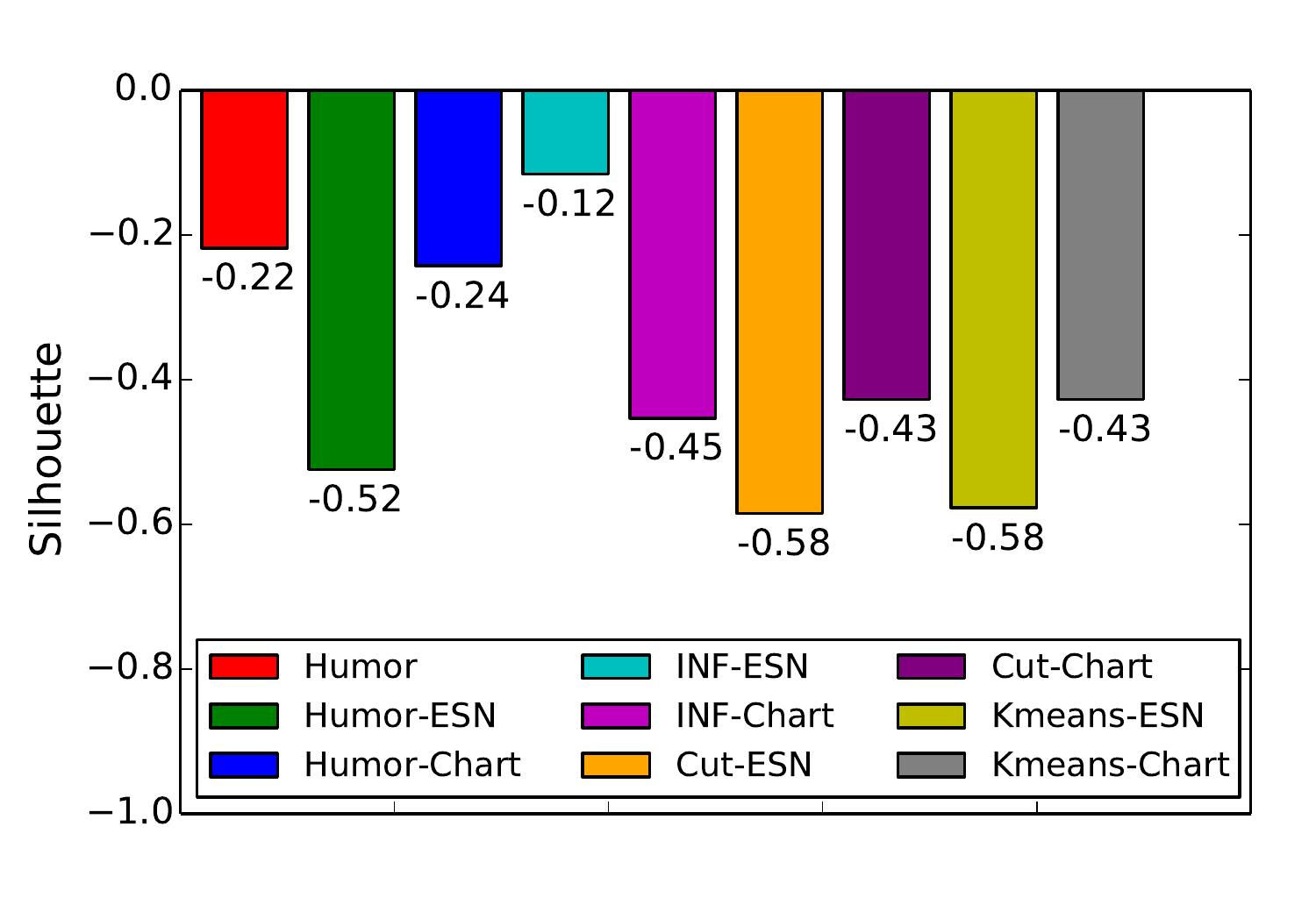}
    \end{minipage}
  }
\subfigure[Normalized-DBI]{ \label{result_3}
    \begin{minipage}[l]{.45\columnwidth}
      \centering
      \includegraphics[width=\textwidth]{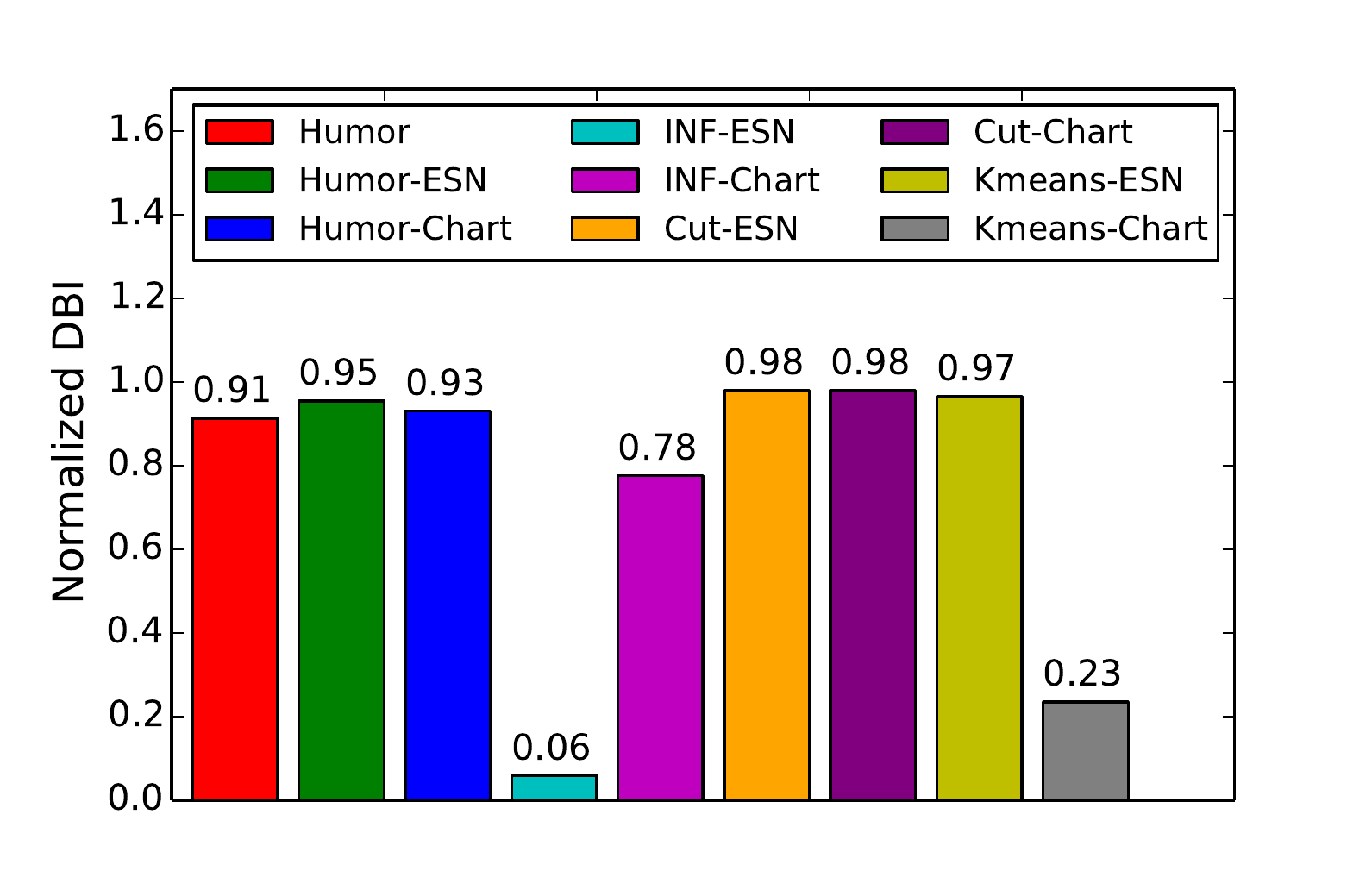}
    \end{minipage}
  }
\subfigure[Entropy]{ \label{result_4}
    \begin{minipage}[l]{.45\columnwidth}
      \centering
      \includegraphics[width=\textwidth]{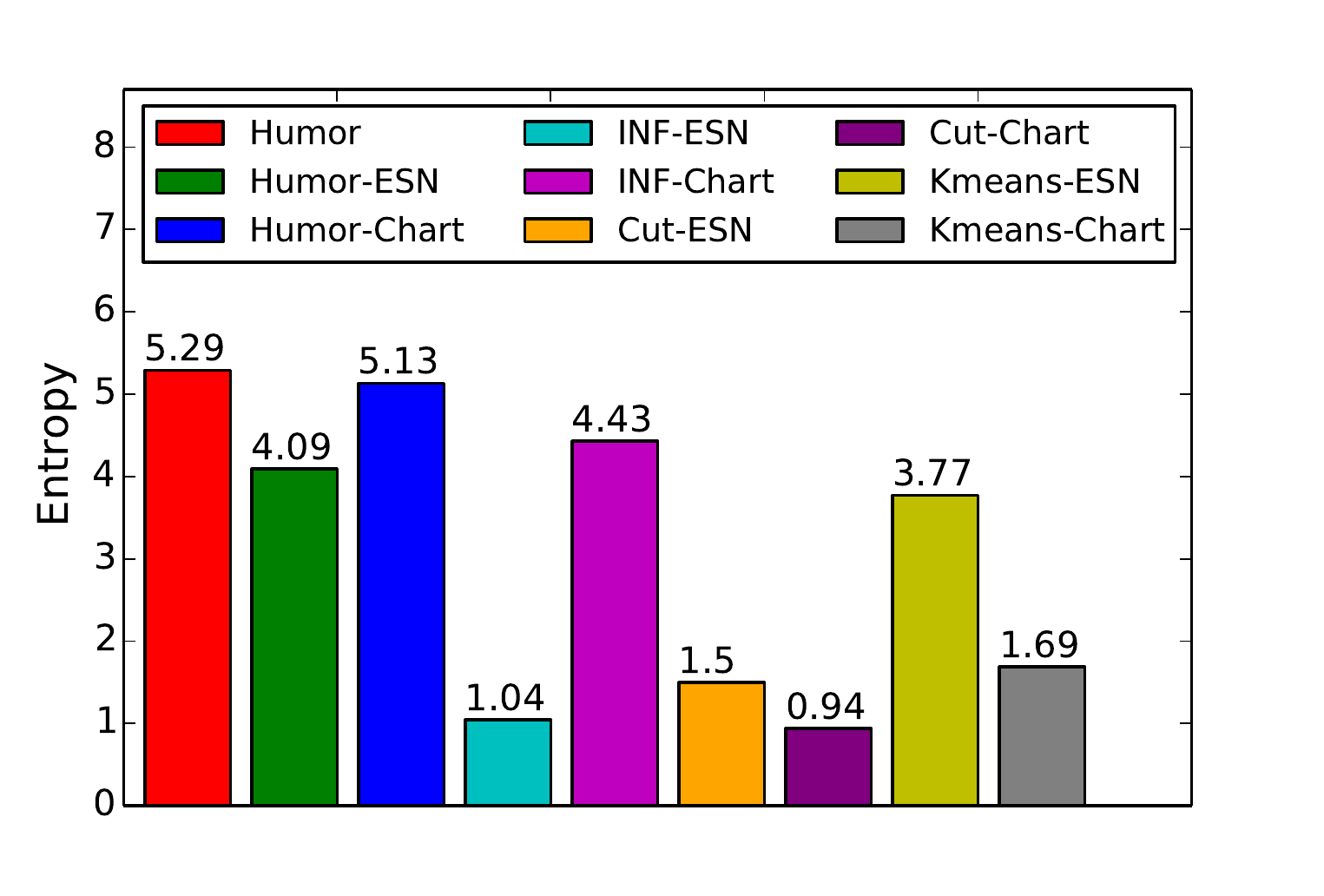}
    \end{minipage}
  }
\vspace{-15pt}
  \caption{Experiment result of comparison methods evaluated by the Density, Silhouette Index, Normalized-DBI and Entropy metrics.}\label{fig:result}\vspace{-15pt}
\end{figure*}

\subsubsection{Comparison Methods}

To show the advantages of the {\our} framework, we compare {\our} with both traditional and state-of-art community detection algorithms, which can be divided into three categories:

\noindent \textbf{Multi-Source Community Detection}

\begin{itemize}

\item \textit{\our}: {\our} is the method proposed in this paper, which utilizes information in both online ESNs and offline company internal information sources to infer the community structure of all the employees in the company. 


\end{itemize}

\noindent \textbf{Single-Source (ESNs Only) Community Detection}

To show that the internal information sources are helpful for providing extra social information about the employees, we apply $4$ other different baseline methods utilizing the information available in the online ESNs only but no information from the company internal information sources.

\begin{itemize}

\item \textit{\ouresn}: We also propose a simplified version of {\our} which just infer the community structure of the employees based on information in online ESNs only.

\item \textit{\infesn}: Method {\infesn} is the influence based clustering method proposed in \cite{ZL13}, which proposes to propagate information among the employees (only the social link information) within one single information source (e.g., online ESNs), where employees with strong mutual influences will be grouped in the same cluster.

\item \textit{\cutesn}: Community detection method ``normalized cut'' \cite{SM00} aims at minimizing the normalized cut of links between different clusters can be used to detect the communities. Method {\cutesn} that minimizes the cut of social links among employees is used as a comparison method in the experiments.

\item \textit{\kmeansesn}: In addition, for completeness, the traditional clustering method KMeans ({\kmeansesn}) is also used as a baseline method, where the social adjacency matrix involving the employees in online ESNs is used as the input of {\kmeansesn}.

\end{itemize}

\noindent \textbf{Single-Source (Company Internal Information Only) Community Detection}

To show that the online ESN is useful in providing extra social information about the employees, we also apply $4$ different baseline methods utilizing the information available in the company internal information sources only.

\begin{itemize}

\item \textit{\ourchart}: {\ourchart} is the simplified version of {\our} corresponding to {\ouresn}, which infers the employees' community structure based on the information available in the company internal information sources. 

\item \textit{\infchart}: Similar to {\infesn}, method {\infchart} builds the influence propagation model based on the organizational chart information among all the employees in the company internal information sources.

\item \textit{\cutchart}: Method {\cutchart} aims at minimizing the cut of management links between employees in the organizational chart in the company internal information sources.

\item \textit{\kmeanschart}: Method {\kmeanschart} uses the adjacency matrix constructed based on the organizational chart as the input and infers the employees community structure.

\end{itemize}

\subsubsection{Evaluation Metrics}

To measure the performance of the comparison methods, different metrics are applied in this paper. 

For the MSR dataset, we have both the predicted community structure $\mathcal{C}_{pred}$ (outputted by the methods) and the real community structure $\mathcal{C}_{true}$ (i.e., the ground truth) about these employees. To evaluate the performance of the methods (i.e., compare $\mathcal{C}_{pred}$ and $\mathcal{C}_{true}$), the metrics used in this paper include \textit{Rand} \cite{NC07}, \textit{Mutual Information} (MI) \cite{NC07}, \textit{Purity} \cite{AGAV09} and \textit{Inverse Purity} \cite{AGAV09}.

In addition, for the community structure $\mathcal{C}_{pred}$ outputted by different comparison methods, we will use $4$ other widely applied metrics \textit{normalized-dbi} \cite{DB79}, \textit{silhouette index} \cite{R87}, \textit{density} \cite{S07}, and \textit{entropy} \cite{NC07} in this paper. Metrics \textit{ndbi}, \textit{silhouette} will use the similarity score among the employees, \textit{density} counts the number of edges in each of the community, and \textit{entropy} measure the distribution of the community sizes. For \textit{ndbi} and \textit{silhouette}, we calculate the average of the $6$ \textit{enterprise intimacy} scores defined in this paper as the real-world similarities among the employees, while for \textit{density}, we count the ratio sum of social links in ESNs and management links in organizational chart as the real-world links among employees.

\subsection{Experiment Results}

According to the experiments, the alternative updating scheme works very well and both the matrix $\mb{U}$ and $\mb{V}$ can converge within 30 iterations. The experiment results obtained based on the converged matrices are available in Figures~\ref{fig:result_truth}-\ref{fig:result}, where plots in Figure~\ref{fig:result_truth} show the results achieved by different methods compared with the ground truth and plots in Figure~\ref{fig:result} show the performance of different methods evaluated by some traditional frequently-used clustering metrics.

According to the plots in Figure~\ref{fig:result_truth}, the enterprise community detection framework {\our} introduced in this paper can perform very well in detecting the real-world community structures. For instance, as shown in Figure~\ref{result_truth_1}, the \textit{Rand} score achieved by {\our} is $0.38$, which denotes among all the potential employee pairs, $38\%$ of them are either correctly grouped in the same community or correctly divided into different communities. The \textit{Rand} score obtained by {\our} is more than $3$ times larger than the scores achieved by methods {\infesn}, {\infchart}, {\cutesn}, {\cutchart}, {\kmeansesn} and {\kmeanschart} respectively. Similar results can be observed when the evaluation metrics are \textit{Mutual Information}, \textit{Purity} and \textit{Inverse Purity} in Figures~\ref{result_truth_2}-\ref{result_truth_4} respectively.

In addition, by comparing {\our} with {\ouresn} and {\ourchart}, based on information from both sources, {\our} can outperform the single-source community detection methods {\ouresn} and {\ourchart} with great advantages. For instance, the \textit{Rand} score achieved by {\our} is $19\%$ higher than the score achieved by {\ourchart}, and over $3$ times greater than the Rand score of {\ouresn}. For other metrics, {\our} and {\ourchart} have comparable performance, and {\our} can outperform {\ouresn} consistently. Meanwhile, according to the results, method {\ourchart} based on the chart performs much better than the method {\ouresn} based on the ESNs. One potential explanation can be that ESNs provide the information for a subset of the total employees in the company only, but the company internal information can cover all the employees in the company. Lots of the employees actually either have no accounts or have no social activities in the ESNs, and the community structure of these users is hard to infer merely based on the information in ESNs. 

In Figure~\ref{fig:result}, we show the results obtained by the comparison methods evaluated by traditional metrics, like \textit{Density}, \textit{Silhouette}, \textit{Normalized-DBI} and \textit{Entropy}. From the results, we can observe that {\our} can perform much better than the other baseline methods when evaluated by \textit{Density}, \textit{Silhouette Index} and \textit{Entropy}, and have comparable performance with some baseline methods when evaluated by \textit{Normalized-DBI}. For instance, the \textit{Density} score obtained by {\our} is $0.74$, which denotes $74\%$ of the following links and management links are preserved within the communities. The density score obtained by {\our} is $3$ times larger than that obtained by {\cutesn} and over $10$ times greater than those obtained by  {\infesn}, {\infchart}, {\cutchart}, {\kmeansesn} and {\kmeanschart}. Similar results can be observed for the \textit{Silhouette Index} and \textit{Entropy} metrics in Figure~\ref{result_2} and Figure~\ref{result_4}.

\begin{figure*}[t]
\vspace{-40pt}
\centering
\subfigure[Density]{ \label{parameter_1}
    \begin{minipage}[l]{.45\columnwidth}
      \centering
      \includegraphics[width=\textwidth]{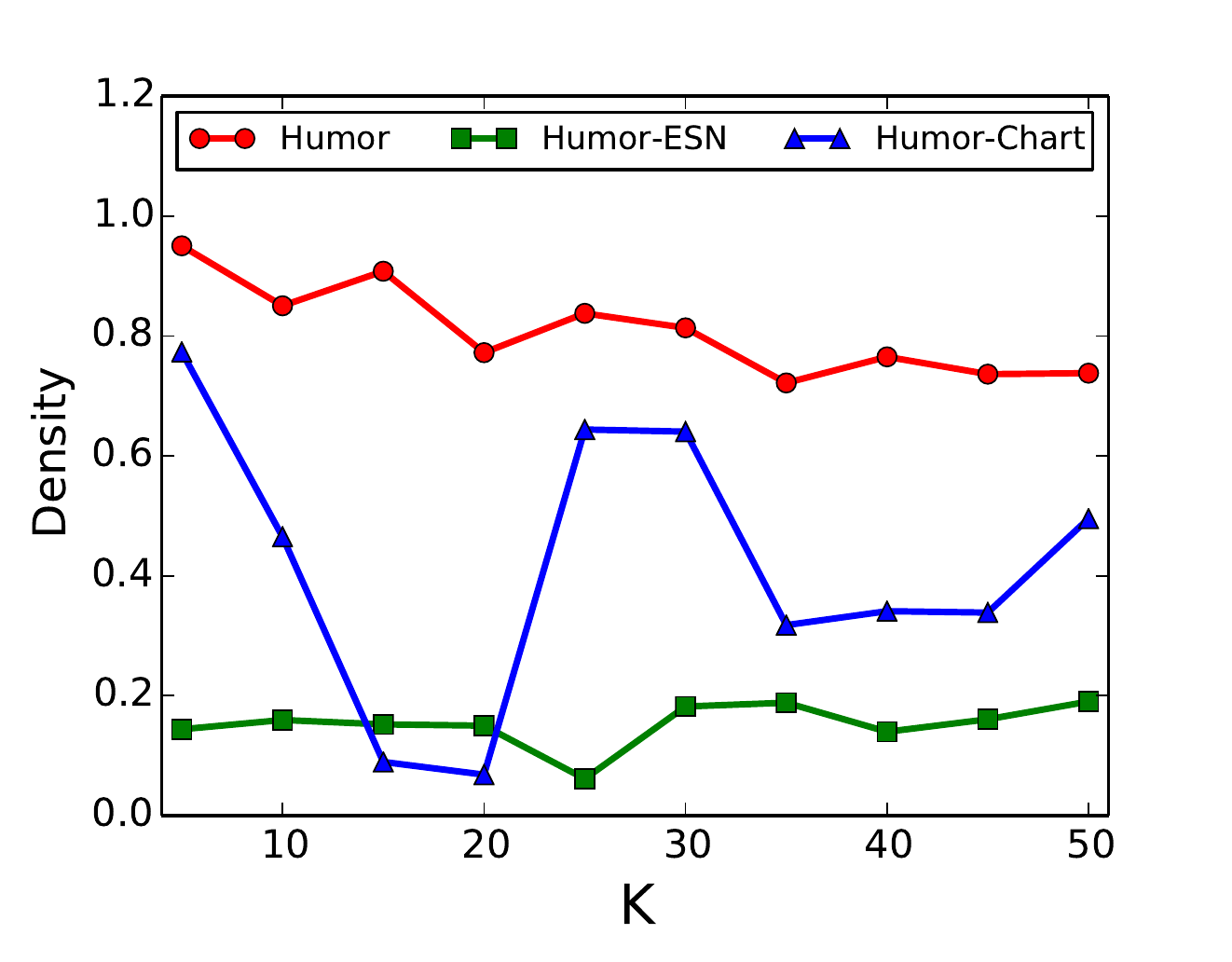}
    \end{minipage}
  }
  \subfigure[Silhouette Index]{\label{parameter_2}
    \begin{minipage}[l]{.45\columnwidth}
      \centering
      \includegraphics[width=\textwidth]{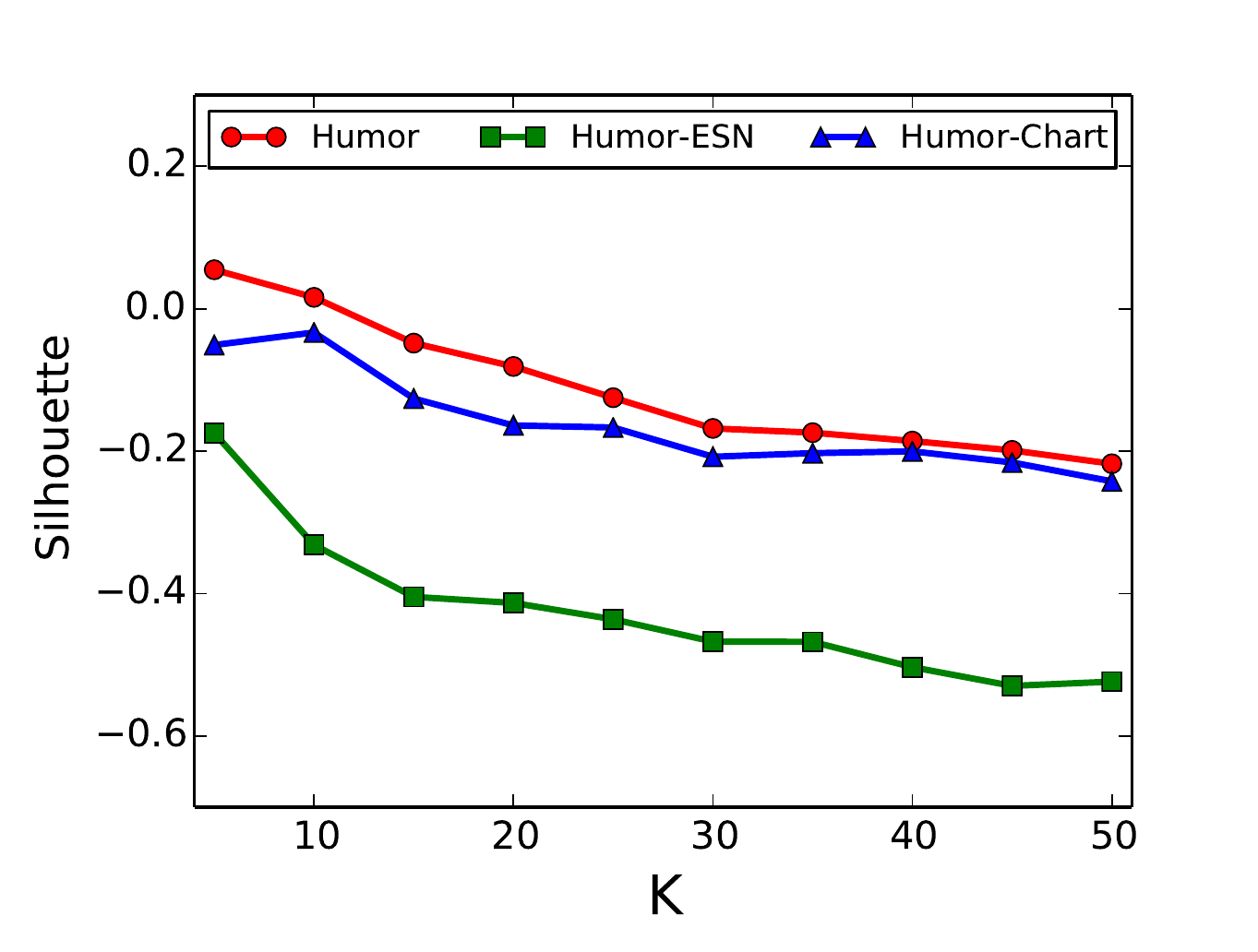}
    \end{minipage}
  }
\subfigure[Normalized-DBI]{ \label{parameter_3}
    \begin{minipage}[l]{.45\columnwidth}
      \centering
      \includegraphics[width=\textwidth]{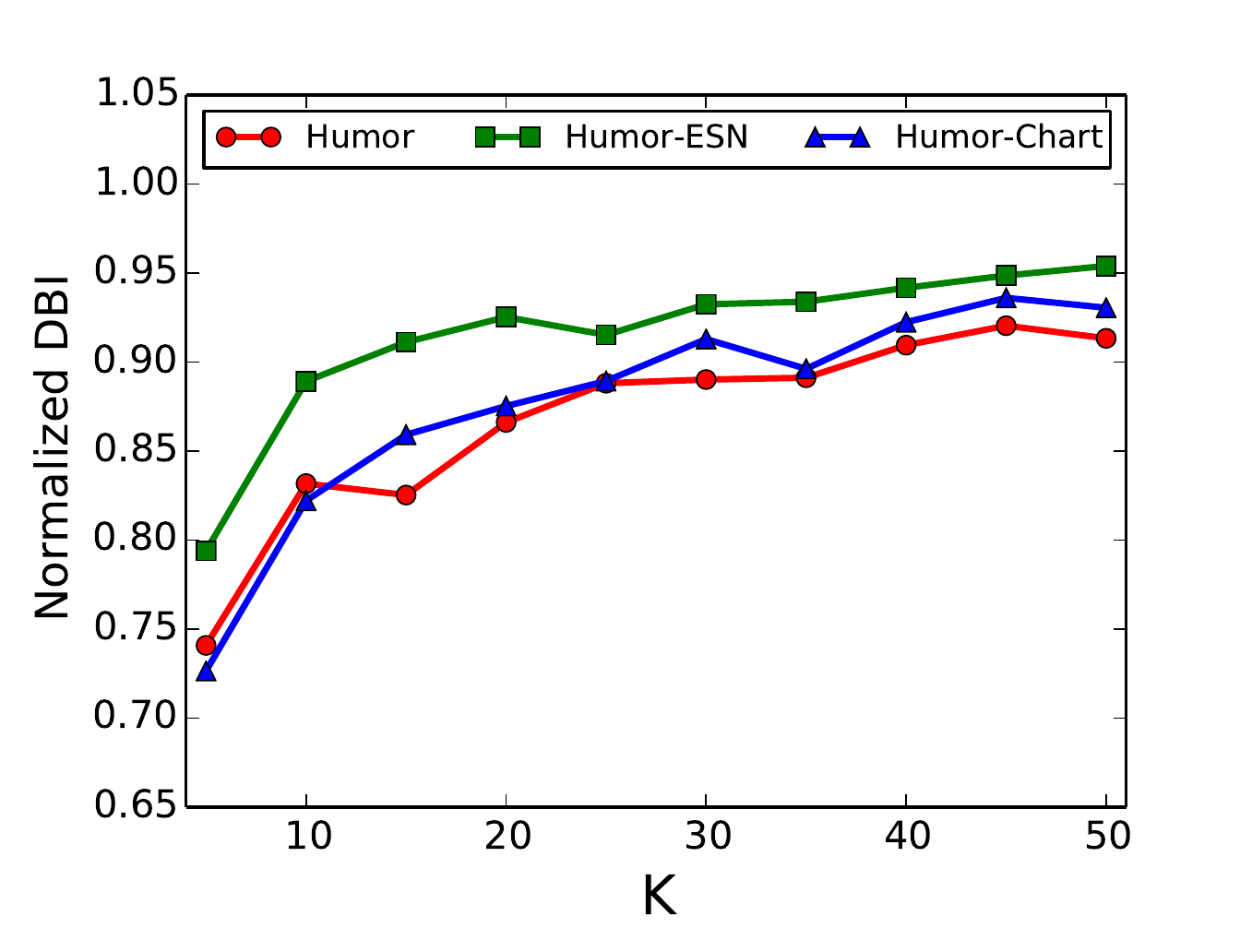}
    \end{minipage}
  }
\subfigure[Entropy]{ \label{parameter_4}
    \begin{minipage}[l]{.45\columnwidth}
      \centering
      \includegraphics[width=\textwidth]{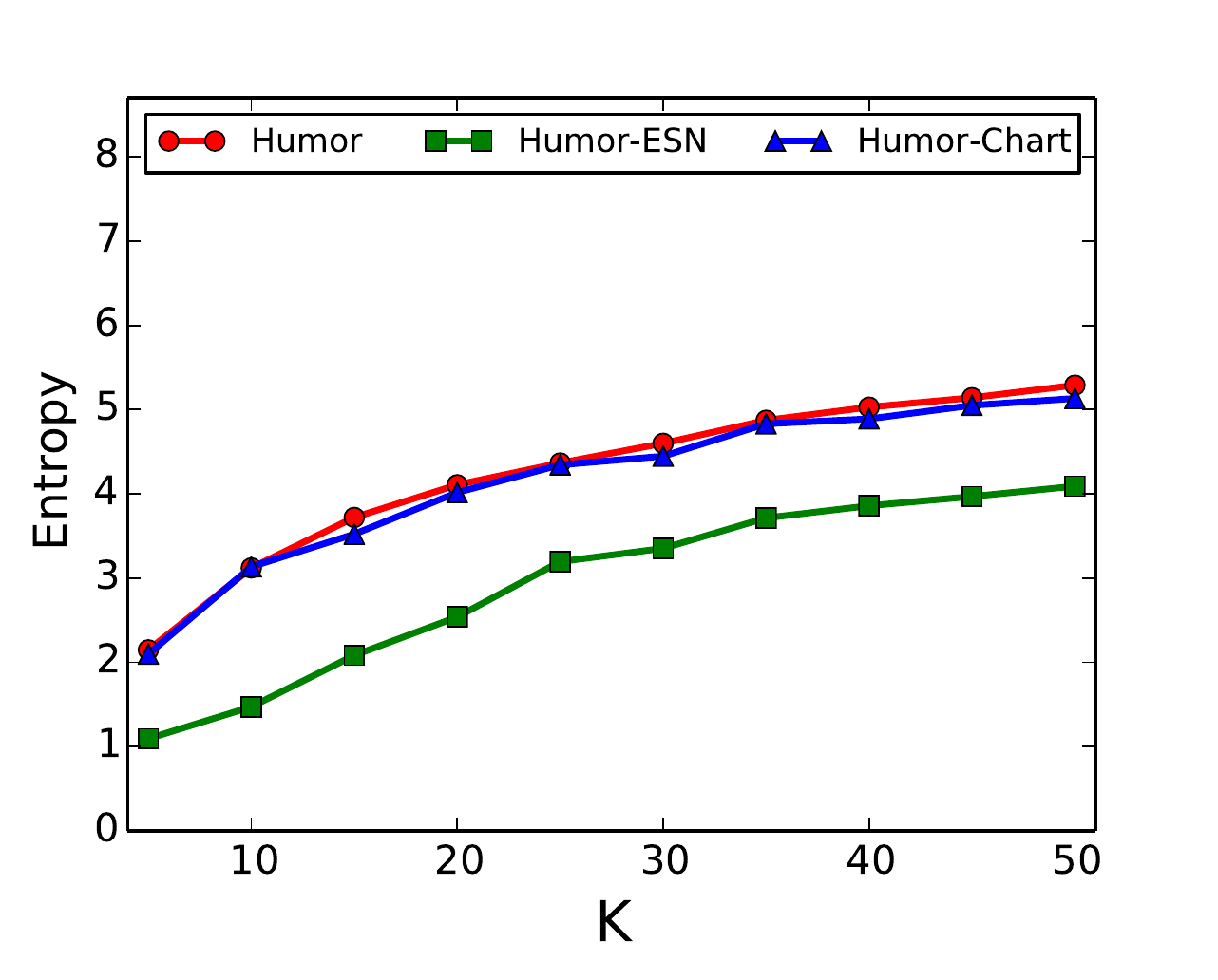}
    \end{minipage}
  }
\vspace{-15pt}
  \caption{Parameter analysis of {\our}, {\ouresn} and {\ourchart} evaluated by the Density, Silhouette Index, Normalized-DBI and Entropy metrics.}\label{fig:result_parameter}\vspace{-15pt}

\end{figure*}

\subsection{Parameter Analysis}

In this part, we will analyze the sensitivity of the community number parameter $K$ on the performance of methods {\our}, {\ouresn} and {\ourchart}. The results evaluated by \textit{Density}, \textit{Silhouette}, \textit{Normalized-DBI} and \textit{Entropy} are available in Figure~\ref{fig:result_parameter}. From the results, we can observe that the performance of these methods depend on parameter $K$ a lot and will change significantly as $K$ varies. For instance, as shown in Figure~\ref{parameter_1} and Figure~\ref{parameter_2}, the \textit{Density} and \textit{Silhouette Index} obtained by these three methods decrease steadily as $K$ increases, but the slopes of the curves tend to be flatter. Similar observations are shown in Figures~\ref{parameter_3}-\ref{parameter_4}. As $K$ increases, the performance of these methods gets better when evaluated by \textit{Normalized-DBI} and \textit{Entropy}, and the curve tend to be more smooth.

\vspace{-3pt}
\section{Related Work} \label{sec:relatedwork}

Clustering aims at grouping similar objects in the same cluster and many different clustering methods have also been proposed. One type is the hierarchical clustering methods \cite{HTF09}, which include agglomerative hierarchical clustering methods \cite{CHS04} and divisive hierarchical clustering methods \cite{CHS04}. Meanwhile, according to the manner that the similarity measure is calculated, the hierarchical clustering methods can be further divided into single-link clustering \cite{SS73}, complete-link clustering \cite{K67} and average-link clustering \cite{W63}. Another type of clustering methods is partition-based methods, which include K-Means for instances with numerical attributes \cite{H98}, K-Medoids for instances with categorical attributes \cite{PJ09}, probabilistic clustering \cite{BBM04}, as well as density-based clustering methods \cite{BR93}. Other clustering methods include grid-based clustering methods \cite{PL04}, constraint-based clustering \cite{TNLH01} and fuzzy clustering \cite{HHK99}.

Clustering method has also been widely used to detect communities in networks. Newman et al. introduce a modularity function measuring the quality of a division of networks \cite{NG04}. Shi et al. introduce the concept of normalized cut and discover that the eigenvectors of the Laplace matrix provide a solution to the normalized cut objective function \cite{SM00}. In addition, many community detection works have been done on heterogeneous online social networks. Sun et al. \cite{SAH12} propose to study the clustering problem with complete link information but incomplete attribute information. Lin et al. \cite{LKYWJL12} try to detect the communities in networks with incomplete relational information but complete attribute information. In recent years, many works have propose to detect communities across aligned networks. Zhang et al. \cite{ZY15} propose to detect the communities in emerging networks with extremely sparse relational information and little attribute information by propagating information from other aligned networks (with abundant of information). Jin et al. \cite{JZYYL14} propose to study the synergistic partition problem of multiple large scale partially aligned social networks based on the Map-Reduce.




Enterprise social networks \cite{ZYL15, ZLY15, ZYLZ16, ZYL17, ZYL17-2} can help employees in companies get reliable information \cite{YMZ12, ELG07}. Yarosh et al. \cite{YMZ12} explore the importance of different information types in expert searching based on a small-sized questionnaire dataset. Based on the heterogeneous information in enterprise social networks, Zhang et al. propose to infer the complete organizational chart based on an unsupervised learning framework CREATE in \cite{ZYL15}. By analyzing the employees' various social behaviors at workplace, Zhang et al. propose to recommend friends for the employees in the online enterprise social networks in \cite{ZLY15}. Workplace has become an important place for social information exchange, and Zhang et al. \cite{ZYLZ16} studies the enterprise information diffusion problem considering the diffusion channels in both online and offline world. Companies can train their employees by involving them into company internal projects, Zhang et al. propose to partition the employees into different teams from the team formation perspective \cite{ZYL17}.

\section{Conclusion}\label{sec:conclusion}

In this paper, we have studied the enterprise community detection problem based on the enterprise information about the employees in both company internal information sources and online ESNs involving the employees. An integrated community detection framework {\our} has been proposed based on the broad learning setting in this paper. {\our} can detect the \textit{micro community structures} about the employees based on each category of the \textit{enterprise intimacy} scores calculated based on one type of enterprise information. In addition, the \textit{micro community structures} about the employees are further fused in {\our} with the \textit{intra-fusion} and \textit{inter-fusion} steps. Extensive experiments have been done on the real-world enterprise datasets (including both the company internal data and the online ESNs data), and the results have demonstrated the outstanding performance of framework {\our}.

\balance
\bibliographystyle{plain}
\bibliography{reference}

\begin{thebibliography}{10}

\bibitem{AGAV09}
E.~Amig\'{o}, J.~Gonzalo, J.~Artiles, and F.~Verdejo.
\newblock A comparison of extrinsic clustering evaluation metrics based on
  formal constraints.
\newblock {\em Information Retrieval}, 2009.

\bibitem{BR93}
J.~Banfield and A.~Raftery.
\newblock Model-based {G}aussian and non-{G}aussian clustering.
\newblock {\em Biometrics}, 1993.

\bibitem{BBM04}
S.~Basu, M.~Bilenko, and R.~Mooney.
\newblock A probabilistic framework for semi-supervised clustering.
\newblock In {\em KDD}, 2004.

\bibitem{CHS04}
P.~Cimiano, A.~Hotho, and S.~Staab.
\newblock Comparing conceptual, divise and agglomerative clustering for
  learning taxonomies from text.
\newblock In {\em ECAI}, 2004.

\bibitem{CW90}
D.~Coppersmith and S.~Winograd.
\newblock Matrix multiplication via arithmetic progressions.
\newblock {\em Journal of Symbolic Computation}, 1990.

\bibitem{DB79}
D.~Davies and D.~Bouldin.
\newblock A cluster separation measure.
\newblock {\em IEEE Transactions on Pattern Analysis and Machine Intelligence},
  1979.

\bibitem{ELG07}
K.~Ehrlich, C.~Lin, and V.~Griffiths-Fisher.
\newblock Searching for experts in the enterprise: Combining text and social
  network analysis.
\newblock In {\em GROUP}, 2007.

\bibitem{HTF09}
T.~Hastie, R.~Tibshirani, and J.~Friedman.
\newblock Hierarchical clustering.
\newblock In {\em The Elements of Statistical Learning}. 2009.

\bibitem{HHK99}
F.~Hopner, F.~Hoppner, and F.~Klawonn.
\newblock {\em Fuzzy Cluster Analysis: Methods for Classification, Data
  Analysis and Image Recognition}.
\newblock 1999.

\bibitem{HXZZGY16}
Q.~Hu, S.~Xie, J.~Zhang, Q.~Zhu, S.~Guo, and P.~Yu.
\newblock Heterosales: Utilizing heterogeneous social networks to identify the
  next enterprise customer.
\newblock In {\em WWW}, 2016.

\bibitem{H98}
Z.~Huang.
\newblock Extensions to the k-means algorithm for clustering large data sets
  with categorical values.
\newblock {\em Data Mining Knowledge Discovery}, 1998.

\bibitem{JZYYL14}
S.~Jin, J.~Zhang, P.~Yu, S.~Yang, and A.~Li.
\newblock Synergistic partitioning in multiple large scale social networks.
\newblock In {\em IEEE BigData}, 2014.

\bibitem{J97}
T.~Joachims.
\newblock A probabilistic analysis of the rocchio algorithm with tfidf for text
  categorization.
\newblock In {\em ICML}, 1997.

\bibitem{K67}
B.~King.
\newblock {Step-Wise Clustering Procedures}.
\newblock {\em Journal of The American Statistical Association}, 1967.

\bibitem{LKYWJL12}
W.~Lin, X.~Kong, P.~Yu, Q.~Wu, Y.~Jia, and C.~Li.
\newblock Community detection in incomplete information networks.
\newblock In {\em WWW}, 2012.

\bibitem{NG04}
M.~Newman and M.~Girvan.
\newblock Finding and evaluating community structure in networks.
\newblock {\em Physical Review}, 2004.

\bibitem{NC07}
N.~Nguyen and R.~Caruana.
\newblock Consensus clusterings.
\newblock In {\em Data Mining, 2007. ICDM 2007. Seventh IEEE International
  Conference on}, 2007.

\bibitem{PJ09}
H.~Park and C.~Jun.
\newblock A simple and fast algorithm for k-medoids clustering.
\newblock {\em Expert Systems with Applications}, 2009.

\bibitem{PL04}
N.~Park and W.~Lee.
\newblock Statistical grid-based clustering over data streams.
\newblock {\em SIGMOD Record}, 2004.

\bibitem{R87}
P.~Rousseeuw.
\newblock Silhouettes: A graphical aid to the interpretation and validation of
  cluster analysis.
\newblock {\em Journal of Computational and Applied Mathematics}, 1987.

\bibitem{S07}
S.~Schaeffer.
\newblock Survey: Graph clustering.
\newblock {\em Computer Science Review}, 2007.

\bibitem{SM00}
J.~Shi and J.~Malik.
\newblock Normalized cuts and image segmentation.
\newblock {\em TPAMI}, 2000.

\bibitem{SS73}
P.~Sneath and R.~Sokal.
\newblock {\em Numerical Taxonomy. The Principles and Practice of Numerical
  Classification}.
\newblock 1973.

\bibitem{SAH12}
Y.~Sun, C.~Aggarwal, and J.~Han.
\newblock Relation strength-aware clustering of heterogeneous information
  networks with incomplete attributes.
\newblock {\em VLDB}, 2012.

\bibitem{TNLH01}
A.~Tung, R.~Ng, L.~Lakshmanan, and J.~Han.
\newblock Constraint-based clustering in large databases.
\newblock In {\em ICDT}, 2001.

\bibitem{W63}
J.~Ward.
\newblock Hierarchical grouping to optimize an objective function.
\newblock {\em Journal of the American Statistical Association}, 1963.

\bibitem{YL13}
J.~Yang and J.~Leskovec.
\newblock Overlapping community detection at scale: A nonnegative matrix
  factorization approach.
\newblock In {\em WSDM}, 2013.

\bibitem{YMZ12}
S.~Yarosh, T.~Matthews, and M.~Zhou.
\newblock Asking the right person: Supporting expertise selection in the
  enterprise.
\newblock In {\em CHI}, 2012.

\bibitem{ZCZCYH17}
J.~Zhang, J.~Chen, S.~Zhi, Y.~Chang, P.~Yu, and J.~Han.
\newblock Link prediction across aligned networks with sparse low rank matrix
  estimation.
\newblock In {\em ICDE}, 2017.

\bibitem{ZLY15}
J.~Zhang, Y.~Lv, and P.~Yu.
\newblock Friend recommendation in enterprise social networks.
\newblock In {\em CIKM}, 2015.

\bibitem{ZY15}
J.~Zhang and P.~Yu.
\newblock Community detection for emerging networks.
\newblock In {\em SDM}, 2015.

\bibitem{ZY15-2}
J.~Zhang and P.~Yu.
\newblock Mcd: Mutual clustering across multiple heterogeneous networks.
\newblock In {\em IEEE BigData Congress}, 2015.

\bibitem{ZY15-4}
J.~Zhang and P.~Yu.
\newblock Multiple anonymized social networks alignment.
\newblock In {\em ICDM}, 2015.

\bibitem{ZYL15}
J.~Zhang, P.~Yu, and Y.~Lv.
\newblock Organizational chart inference.
\newblock In {\em KDD}, 2015.

\bibitem{ZYL17}
J~Zhang, P.~Yu, and Y.~Lv.
\newblock Enterprise community detection.
\newblock In {\em ICDE}, 2017.

\bibitem{ZYL17-2}
J.~Zhang, P.~Yu, and Y.~Lv.
\newblock Enterprise employee training via project team formation.
\newblock In {\em WSDM}, 2017.

\bibitem{ZYLZ16}
J.~Zhang, P.~Yu, Y.~Lv, and Q.~Zhan.
\newblock Information diffusion at workplace.
\newblock In {\em CIKM}, 2016.

\bibitem{ZYZ14}
J.~Zhang, P.~Yu, and Z.~Zhou.
\newblock Meta-path based multi-network collective link prediction.
\newblock In {\em KDD}, 2014.

\bibitem{ZL13}
Y.~Zhou and L.~Liu.
\newblock Social influence based clustering of heterogeneous information
  networks.
\newblock In {\em KDD}, 2013.

\bibitem{ZZHWZZY17}
J.~Zhu, J.~Zhang, L.~He, Q.~Wu, Y.~Jia, B.~Zhou, C.~Zhang, and P.~Yu.
\newblock Broad learning based multi-source collaborative recommendation.
\newblock In {\em CIKM}, 2017.

\end{thebibliography}

\end{document}